\documentclass{scrartcl}
\usepackage[utf8]{inputenc}
\usepackage{amsmath,amssymb,amsthm,dsfont,mathtools}
\usepackage{enumitem}
\usepackage{graphicx}
\usepackage{tcolorbox}
\usepackage{url}
\usepackage{color}
\usepackage[backref, 					
	bookmarksnumbered,
	pagebackref=true, 
	hyperfootnotes=true, 
	linktoc = page, 
	colorlinks = true, 
	linkcolor = blue,
	citecolor = blue]{hyperref}
\usepackage[capitalise,sort]{cleveref} 

\usepackage[realmainfile]{currfile}
\usepackage{mathtools}

\usepackage{booktabs}
\usepackage{aliascnt}

\usepackage{caption}
\usepackage{pgfplots}
\pgfplotsset{width=10cm,compat=1.9}
\usepgfplotslibrary{external}

\theoremstyle{plain}
\newtheorem{theorem}{Theorem}

\numberwithin{theorem}{section}
\numberwithin{equation}{section}

\newaliascnt{proposition}{theorem}

\aliascntresetthe{proposition}

\newaliascnt{lemma}{theorem}
\newtheorem{lemma}[lemma]{Lemma}
\aliascntresetthe{lemma}

\newaliascnt{corollary}{theorem}
\newtheorem{corollary}[corollary]{Corollary}
\aliascntresetthe{corollary}

\newaliascnt{conjecture}{theorem}

\aliascntresetthe{conjecture}

\theoremstyle{definition}
\newaliascnt{definition}{theorem}
\newtheorem{definition}[definition]{Definition}
\aliascntresetthe{definition}

\newaliascnt{example}{theorem}
\newtheorem{example}[example]{Example}
\aliascntresetthe{example}

\newaliascnt{remark}{theorem}
\newtheorem{remark}[remark]{Remark}
\aliascntresetthe{remark}

\theoremstyle{remark}	

\Crefname{step}{Step}{Steps}
\makeatletter
\@addtoreset{step}{theorem}
\makeatother

\newtheorem*{example*}{Example}

\newaliascnt{assumption}{theorem}
\newtheorem{assumption}[assumption]{Assumption}
\tcolorboxenvironment{assumption}{
  sharp corners,
  boxrule=0.4pt,
  colback=white,
  before skip=\topsep,
  after skip=\topsep,
}
\aliascntresetthe{assumption}

\crefname{theorem}{Theorem}{Theorems}
\crefname{proposition}{Proposition}{Propositions}
\crefname{lemma}{Lemma}{Lemmas}
\crefname{corollary}{Corollary}{Corollaries}
\crefname{definition}{Definition}{Definitions}
\crefname{example}{Example}{Examples}
\crefname{remark}{Remark}{Remarks}
\crefname{assumption}{Assumption}{Assumptions}

\DeclareMathOperator{\VaR}{VaR}

\DeclareMathOperator{\CCR}{CCR}

\DeclareMathOperator{\ICCR}{ICCR}

\DeclareMathOperator{\TVaR}{TVaR}

\title{The effect of policy cancellation on the risk of an insurance portfolio}

\author{
    Stefan Gerhold\\
    TU Wien\\
    \texttt{sgerhold@fam.tuwien.ac.at}
    \and
    Friedrich Hubalek\\
    TU Wien\\
    \texttt{fhubalek@fam.tuwien.ac.at}
    \and
    Manuel Schranzhofer\\
    TU Wien\\ UNIQA Österreich Versicherungen AG\\
    \texttt{manuel.schranzhofer@tuwien.ac.at}
}

\date{November 18, 2026}

\begin{document}

\maketitle

\begin{abstract}

For an insurance portfolio modelled by a large number of i.i.d.\ losses, the average loss converges to its expected value by the law of large numbers, reflecting the diversification effect of pooling risks. Policy cancellations reduce the portfolio size and thereby weaken this effect. We quantify the resulting increase in risk by applying risk measures to the average loss before and after the cancellation of policies.
For a portfolio of size $n$, we show that the impact of cancelling a single contract is of order $n^{-3/2}$, while the cancellation of a fixed proportion of contracts has an effect of order $n^{-1/2}$. We further extend the proportional cancellation regime to heterogeneous portfolios with different cancellation rates and risk characteristics across segments.
These asymptotic results provide a quantitative basis for assessing surrender fees and cancellation costs from the insurer’s perspective. We complement the theoretical analysis with numerical results for representative life insurance products, illustrating the magnitude of cancellation costs and the effect of portfolio aggregation.\par

The two regimes require different techniques. For the cancellation of a fixed proportion of contracts, we employ central limit theorems in Wasserstein distance. For the cancellation of a single contract, we combine Kantorovich--Rubinstein duality with higher-order distributional approximations based on the Edgeworth and Cornish--Fisher expansions.

\end{abstract}



\section{Introduction}

It is a cornerstone of the insurance business that risk sharing works best if the number of contracts is very large. Vice versa, early termination of a contract weakens diversification and can cause additional costs for the insurer, which justify a surrender fee.
While our investigation of this fee might be of relevance in non-life insurance as well, and possibly for prepayment charges for a credit portfolio, we focus on life insurance for motivating and interpreting our findings.
Modelling life insurance contracts with a \emph{given} surrender charge is standard; see e.g.~\cite{ChDeSt22} and the references therein, and Section~VI in~\cite{Li05} for an early reference.
For a recent analysis of cancellations from the viewpoint of behavioural economics, we refer to~\cite{GoSm21}.
Reasons to discourage cancellations, and apply surrender charges -- if legally applicable -- include, among others: negative risk selection, failure to properly contribute to solvency requirements, cost of asset liquidation, unfavourable yield curve movements, and duration mismatch. While these have been considered in several works~(e.g.\ \cite{DAV22,DOSo16,Burkhart2018,Hacariz2026,Chang2022}),  the probabilistic effect that the fluctuation risk increases if the portfolio shrinks  has not been analysed mathematically, and we aim to fill this gap in the actuarial literature.

\pagebreak[2]
Throughout this article, $(X_k)_{k\in\mathbb{N}}$ is an i.i.d.\ sequence of  real random variables, modelling insurance losses, i.e.\ discounted future benefits minus discounted future premia.
From the insurer's point of view, the distribution of the average loss $\bar{X}_{n}\coloneqq\frac1n \sum_{k=1}^n X_k$ seems preferable to that of $\bar{X}_{n-1}$, and it is our aim to give rigorous results on this.
To quantify the effect, we apply a risk measure to the average loss before and after policy cancellations and study the difference asymptotically as the portfolio size~$n$ tends to infinity. Our main contribution is to derive explicit asymptotics  under different cancellation regimes. We show that for a large class of risk measures --- including VaR and TVaR --- the impact of cancelling a single contract is of order $n^{-3/2}$, whereas the cancellation of a fixed \emph{proportion} of contracts leads to order $n^{-1/2}$. In both cases,
the cancellation cost per cancelling policyholder is of order $n^{-1/2}$. The following simple approximation of this cost is one of our main results:
\begin{equation}\label{eq:ccr intro}
  \CCR_{\theta,n}\approx
  (\tfrac{1}{2}-\tfrac{1}{8} \theta) \sigma \rho(\Phi)\, n^{-1/2}, \quad n\to\infty.
\end{equation}
The ``R'' in the notation CCR stresses that increased average risk is the only driver of the cancellation cost that we consider. How to properly
aggregate this with the sources mentioned above is a natural topic for future research.
The expression~\eqref{eq:ccr intro} makes transparent how the cancellation cost depends on the portfolio size~$n$, the cancellation proportion~$\theta$, the standard
deviation~$\sigma$ of the individual losses, and the chosen risk measure~$\rho$ through its value at the standard Gaussian distribution. We will make this relationship precise for a broad class of loss distributions and risk measures.

\pagebreak[2]
Explicit expressions for the corresponding leading-order coefficients are provided in \cref{thm:::AsymptoticsConstantVaR} and \cref{thm:::AsymptoticConstantLp} as well as \cref{lem:::AsymptoticProportionalVaR} and \cref{thm:::AsymptoticProportionalLp}. 
We further extend the proportional cancellation regime to heterogeneous portfolios, allowing for different cancellation rates and risk characteristics across portfolio segments (see \cref{lem:HeterogeneousProportionalCancellation}). In particular, we show that portfolio diversification does not necessarily reduce the cancellation cost for individual segments, as the effect depends on the cancellation rates and risk characteristics of the other segments in the portfolio (see \cref{exm:::FailureAggregation}). Finally, we show that the first-order asymptotics in the proportional cancellation regime remain unchanged when cancellations are modelled stochastically by independent Bernoulli random variables, provided that the cancellations are independent of the losses (see \cref{thm:::StochasticExtensionOfProportionalCancellation}).

\pagebreak[2]
The remainder of the paper is organised as follows. In \cref{sec:::Preliminaries}, we introduce the basic notation and discuss Gaussian losses, which serve as a blueprint for the subsequent analysis. We also discuss the scope and limitations of our approach, including the class of risk measures considered.
In \cref{sec:::ConstantCancellationRegime}, we study the constant cancellation regime, where a single policy, or more generally a fixed number of policies, is cancelled.
The analysis relies on Edgeworth expansions and Kantorovich--Rubinstein duality to derive first-order asymptotics of the cancellation cost. While this seems to be a natural first step to understand the change in individual risk contribution, readers interested in actuarial applications may wish to skip to \Cref{sec:::ProportionalCancellationRegime}. It is devoted to the proportional cancellation regime, which is more relevant in practice, and mathematically simpler. Here, the main analytical tool is a Central Limit Theorem in Wasserstein distance, which allows us to establish asymptotic results for a broad class of law-invariant risk measures. 
We then consider heterogeneous portfolios and stochastic cancellation in Subsections~\ref{subsec:::ProportionalCancellationInheterogeneousPortfolio} and \ref{sec:::ExtensionStochasticCancellation}, respectively.
Finally, \cref{sec:::NumericalResults} illustrates the theoretical results numerically, for term life and pure endowment insurance contracts. Within a general Markov-chain formulation, we derive a recursive procedure for calculating the variance of future cash flows, which is required to evaluate the approximation of the cancellation cost $\mathrm{CCR}$. The numerical examples consider both standalone new-business cohorts and contracts within an existing portfolio, illustrating our results for concrete life insurance products and the effect of portfolio aggregation across cohorts with different remaining maturities.

\pagebreak[2]
\section{Preliminaries}\label{se:prelim}
\label{sec:::Preliminaries}

\subsection{Cancellation cost and notation}
\label{sec:::CancellationCostAndNotation}

If a risk measure~$\rho$ is fixed, then the risk of the portfolio is~$\rho\big(\sum_{k=1}^n X_k\big)$ before cancellation, and~$\rho\big(\sum_{k=1}^{n-m} X_k\big)$ after cancellation of~$m$ contracts. From the perspective of an individual policyholder, the collective loss allocated to him/her is given by the average loss per policyholder,
\begin{equation*}
\bar X_n = \frac{1}{n}\sum_{k=1}^n X_k
\quad\text{and}\quad
\bar X_{n-m} = \frac{1}{n-m}\sum_{k=1}^{n-m} X_k,
\end{equation*}
respectively. Thus, $\bar X_n$ represents the loss allocated to an individual policyholder through the pooling and sharing of risks within the collective. A cancellation changes the composition and size of this collective, and may therefore alter the risk allocated to each remaining policyholder. The resulting change in this individual risk contribution is an important quantity considered throughout the paper.
The standard definitions concerning risk measures that we use are recalled
in Subsection~\ref{se:risk measures} below.

\begin{definition}[Change in individual risk contribution]
\label{def:CCR}
Let $\rho$ be a law-invariant cash-invariant and positively homogeneous risk measure and $(X_k)_{k\in\mathbb{N}}$ be a sequence of i.i.d.\ copies of a random variable $X$. For $n \geq m \geq 0$, we define
\begin{equation}\label{eq:def Delta}
\Delta_{m,n}(X,\rho)
\coloneqq
\rho(\bar X_{n-m})-\rho(\bar X_n),
\end{equation}
with the convention $\bar X_0 \equiv 0$. We call $\Delta_{m,n}(X,\rho)$ the \textit{change in individual risk contribution}. If the risk measure $\rho$ is clear from the context, we simply write $\Delta_{m,n}(X)$, and if the underlying sequence is clear as well, we write $\Delta_{m,n}$.
\end{definition}

We proceed to the monetary significance of $\Delta_{m,n}$ and consider
the case $n>m\geq 1$. If the insurer wants to maintain the same risk level per policyholder after cancellation, the amount of cash that has to be added to the remaining portfolio -- i.e., subtracted from
the total loss -- is
\begin{align*}
\ICCR_{m,n} &\coloneqq 
\inf\biggl\{c\in\mathbb{R} \,\bigg|\,
\frac{1}{n}\rho\bigg(\sum_{k=1}^n X_k\bigg)
\geq
\frac{1}{n-m}
\rho\bigg(\sum_{k=1}^{n-m}X_k-c\bigg)\biggr\} \\
&= \inf\biggl\{c\in\mathbb{R} \,\bigg|\,
\rho(\bar{X}_n)
\geq
\rho(\bar{X}_{n-m}) - \frac{c}{n-m}\biggr\}\\
&=(n-m)\Delta_{m,n}(X,\rho),
\end{align*}
where we have used cash-invariance and positive homogeneity. Thus, ICCR is the total compensation required to offset the increase in average risk for the remaining policyholders. The corresponding cost per cancelling policyholder is
\begin{equation}\label{eq:CCR etc}
\CCR_{m,n}(X,\rho)
\coloneqq
\frac{1}{m} \ICCR_{m,n}(X,\rho) = \frac{n-m}{m} \Delta_{m,n}(X,\rho).
\end{equation}
The notation $\CCR_{\theta,n}$ from~\eqref{eq:ccr intro}, with $0<\theta<1$ the cancellation proportion, is defined in~\cref{sec:::DynmaicsOfCancellation}.

\begin{remark}
  We will state most of our results for the change in individual risk contribution
  $\Delta_{m,n}$, as they easily translate to the cancellation cost
  by~\eqref{eq:CCR etc}.
\end{remark}

\begin{remark}
We observe that the loss variables may be assumed to be standardised without loss of generality. Indeed, if~$X_k=\mu+\sigma Z_k$ for~$k\geq1$ with~$\mu\in\mathbb R$ and~$\sigma> 0$, then
\begin{align}\label{eq:normalize}
\begin{split}
\Delta_{m,n}(X,\rho)&=\rho(\bar X_{n-m})-\rho(\bar X_n) \\
&=\sigma\big(\rho(\bar Z_{n-m})-\rho(\bar Z_n)\big)=\sigma\Delta_{m,n}(Z_1; \rho).
\end{split}
\end{align}
\end{remark}

\begin{remark}
It is not obvious whether $\Delta_{m,n}$ is non-negative. Below, we give several asymptotic results that ensure non-negativity for~$n$ sufficiently large. For a non-asymptotic result, note that~$\bar X_n$ decreases with respect to convex order (\cite{ArVi86}, \cite[Corollary~1.5.24]{MuSt02}), and use Theorem~4.3 in~\cite{BaMu06}.
An example with negative $\Delta_{m,n}$, of little relevance to life insurance though, is given
in Example~\ref{ex:stable}.
\end{remark}

\emph{Further notation:} $\Phi$ denotes the cdf and $\phi$ the pdf of the standard normal distribution,
$F^\leftarrow$ the generalized inverse of the cdf~$F$ (see~\eqref{eq:def VaR}),
$F_n$ the cdf of $\sqrt{n}\bar{X}_n$, $\mathcal{M}_1$ the set of probability measures and $\mathcal{M}_1^p$ the set of probability measures with finite $p$-th moment, $W_p$ the Wasserstein distance (see~\eqref{eq:::WassersteinMetric}), and $\Pi(P,Q)$ the set of probability
measures with marginals $P,Q$.

\pagebreak[2]
\subsection{Cancellation regimes and normally distributed loss}
\label{sec:::DynmaicsOfCancellation}

There are different ways to model the relationship between the number of cancellations and the portfolio size. We first consider deterministic cancellation regimes. 

\begin{enumerate}[label=(\alph*)]
\item \textit{Constant cancellation:}
We fix the number of cancellations to $m \in \mathbb{N}$ and vary only the portfolio size $n$.

\item\label{DynamicsOfCancellation:::ProportionalCancellation} \textit{Proportional cancellation:}
We assume that the number of cancellations is directly proportional to the portfolio size. For a constant factor $\theta \in (0,1)$, we define
\begin{equation*}
\Delta_{\theta,n} \coloneqq \Delta_{\lfloor \theta n \rfloor, n}, \quad n \ge 0,
\end{equation*}
and use analogous notation  $\CCR_{\theta,n}$ for the cancellation cost.
\end{enumerate}

Later, in \cref{sec:::ExtensionStochasticCancellation}, we extend our results to a stochastic cancellation regime. A formal calculation based on the CLT, with~$\sigma$ the standard deviation of~$X$, yields
\begin{align*}
  \rho(\bar{X}_{n-m}) &- \rho(\bar{X}_{n})
  = \frac{\sigma}{\sqrt{n-m}} \rho\Big( \frac{\sqrt{n-m}}{\sigma}\bar{X}_{n-m}\Big)
    - \frac{\sigma}{\sqrt{n}} \rho\Big( \frac{\sqrt{n}}{\sigma}\bar{X}_{n}\Big) \\
&\approx \sigma \bigg(\frac{1}{\sqrt{n-m}} - \frac{1}{\sqrt n}\bigg) \rho(\Phi)
  \sim 
  \begin{cases}
    \frac{m\sigma  \rho(\Phi)}{2n^{3/2}}, & m\geq1 \text{ fixed}, \\
    \frac{1-(1-\theta)^{1/2}}{(1-\theta)^{1/2}}
\frac{\sigma\rho(\Phi)}{n^{1/2}}, & m=\lfloor \theta n \rfloor.
  \end{cases}
\end{align*}
Much of what follows is about making the approximation
``$\approx$'' rigorous.
Things are easy if~$X$ is Gaussian, since the cancellation cost admits an explicit representation. Moreover, the asymptotic behaviour derived in the following example will later be extended to more general loss distributions by suitable approximation arguments.

\pagebreak[2]
\begin{example}[Standard normal losses]\label{exm:::NormalDistribution}
Suppose that the losses are standardised and Gaussian, i.e.\ $X \sim N(0,1)$. Let $\rho$ be a positively homogeneous law-invariant risk measure satisfying $\rho(X)<\infty$. Since $\bar X_n \sim N(0,n^{-1})$, positive homogeneity yields
\begin{equation}
\label{eq:::NormalDistributionCancellationCost}
\Delta_{m,n}(\Phi,\rho) = 
\biggl( \frac{1}{\sqrt{n-m}}-\frac{1}{\sqrt{n}}\biggr) \rho(\Phi).
\end{equation}
where $\Phi$ denotes the standard normal cdf. We consider the constant and proportional cancellation regimes separately:
\begin{enumerate}[label=(\alph*)]
    
\item\textit{Constant cancellation:} For fixed $m \in \mathbb{N}$, expanding $\frac{1}{\sqrt{n-m}}-\frac{1}{\sqrt{n}}$ using the binomial series gives
\begin{equation}\label{eq:constant Delta}
\Delta_{m,n}(\Phi,\rho) = \frac{m}{2n^{3/2}}\rho(\Phi)  + O\biggl( \frac{1}{n^{5/2}}\biggr),
\quad n\to\infty.
\end{equation}

\item\textit{Proportional cancellation:} 
As~$\lfloor \theta n \rfloor=  \theta n + O(1)$, the binomial series gives $(n-\lfloor \theta n \rfloor)^{-1/2}=n^{-1/2}(1-\theta)^{-1/2} + O(n^{-3/2})$.
Therefore, substituting~$m=\lfloor\theta n\rfloor$
into~\eqref{eq:::NormalDistributionCancellationCost} yields
\begin{equation}\label{eq:prop Delta}
\Delta_{\theta,n}(\Phi,\rho) =  \frac{1-(1-\theta)^{1/2}}{(1-\theta)^{1/2}}
\frac{\rho(\Phi)}{n^{1/2}} + O\biggl(\frac{1}{n^{3/2}}\biggr).
\end{equation}
\end{enumerate}
\end{example}

\pagebreak[2]
The leading terms in~\eqref{eq:constant Delta} and~\eqref{eq:prop Delta} are the asymptotic orders $n^{-3/2}$ resp.\ $n^{-1/2}$ that we mentioned before~\eqref{eq:ccr intro}.
Consequently, in the proportional cancellation regime, the insurer-level cost satisfies
\begin{equation*}
\ICCR_{\theta,n}(\Phi,\rho)
=
\lambda_\theta \rho(\Phi)n^{1/2}
+
O(n^{-1/2}),
\quad\text{where }
\lambda_\theta
=
\sqrt{1-\theta}-(1-\theta),
\end{equation*}
while the corresponding cost per cancelling policyholder is
\begin{equation*}
\CCR_{\theta,n}(\Phi,\rho) = \frac{\lambda_\theta}{\theta} \rho(\Phi) n^{-1/2} + O(n^{-3/2}).
\end{equation*}
A first-order Taylor expansion of $\lambda_\theta/\theta$ around $\theta=0$ gives $\lambda_\theta/\theta = \tfrac{1}{2}-\tfrac{1}{8}\theta + O(\theta^2)$, yielding the approximation~\eqref{eq:ccr intro}. See Section~\ref{sec:::ProportionalCancellationRegime} for corresponding results for non-Gaussian distributions.
The coefficient~$\lambda_\theta$ is maximized at~$\theta=3/4$, where~$\lambda_{3/4}=1/4$. Thus, in the Gaussian benchmark, a cancellation proportion of $75\%$ yields the largest asymptotic insurer-level cancellation cost. The leading-order coefficient vanishes both as~$\theta\to0$ and as~$\theta\to1$. This can be explained heuristically by two opposing effects. For small~$\theta$, only a small proportion of the portfolio is cancelled, so the resulting reduction in diversification has a limited impact. In contrast, as~$\theta\to1$, only a small proportion of the portfolio remains, leaving few contracts over which the increase in average risk can translate into an insurer-level cost. These two effects suggest that the largest impact should occur at an intermediate cancellation proportion, as confirmed by the maximum at~$\theta=3/4$.

\pagebreak[3]
\subsection{Risk measures}\label{se:risk measures}

Our goal is to extend the asymptotic results obtained for Gaussian losses to more general loss distributions for a broad class of risk measures. We use the loss convention for risk measures, under which larger values represent larger losses. With this convention, we recall the definition of coherent risk measures.

\pagebreak[2]
\begin{definition}[{Risk measures \cite{Artzner1999}}]
\label{def:::RiskMeasure}

Let $\mathcal{X}$ be a vector space of random variables on a probability space $(\Omega,\mathcal{F},\mathbb{P})$. A \textup{risk measure} is a functional $\rho \colon \mathcal{X} \to (-\infty,\infty]$. It is called \textup{proper} if it is not identically $\infty$ and $\rho(X_1) = \rho(X_2)$ for all $X_1 \overset{\mathrm{a.s.}}{=}X_2$ in $\mathcal{X}$. We call a proper risk measure \textup{monetary}, if it satisfies the following properties:
\pagebreak[2]
\begin{enumerate}[label=(\alph*)]
\item\label{itm:::Monotonicity} \textup{Monotonicity:} $\rho(X_1) \le \rho(X_2)$ for all $X_1,X_2 \in \mathcal{X}$ with $X_1 \le X_2$.
\item\label{itm:::CashInvariant} \textup{Cash-invariance:} $\rho(X+m) = \rho(X)+m$ for all $X \in \mathcal{X}$ and $m\in\mathbb{R}$.
\end{enumerate}
\pagebreak[2]
A proper risk measure is called \textup{coherent} if it is monetary
(i.e.~\ref{itm:::Monotonicity} and~\ref{itm:::CashInvariant} hold) and it additionally satisfies the following properties:
\begin{enumerate}[resume, label=(\alph*)]
\item\label{itm:::PosHomo} \textup{Positive homogeneity:} $\rho(\lambda X) = \lambda \rho(X)$ for all $X \in \mathcal{X}$ and $\lambda \in (0,\infty)$.
\item\label{itm:::Subadd} \textup{Subadditivity:} $\rho(X_1+X_2) \le \rho(X_1) + \rho(X_2)$ for all $X_1,X_2 \in \mathcal{X}$.
\end{enumerate}
Further, we call a proper risk measure \textup{law-invariant} if~$\rho(X_1) = \rho(X_2)$ for all~$X_1,X_2 \in \mathcal{X}$ with equal law.
\end{definition}

\pagebreak[2]
\begin{example}[Essential supremum as risk measure]
\label{exm:::EssentialSupremum}

Assume that we are given an i.i.d.\ sequence of bounded random variables $(X_k)_{k\in\mathbb{N}}$ and we consider the coherent risk measure $\rho = \| \cdot \|_\infty$. Then
\begin{equation*}
\|\bar X_n \|_\infty = 
\|X_1\|_\infty, \quad n \in \mathbb{N},
\end{equation*}
which implies $\Delta_{m,n}(X,\rho) = 0$.
\end{example}

\pagebreak[2]
The preceding example demonstrates that law-invariance and coherence alone are insufficient to guarantee the asymptotic behaviour we found for Gaussian losses. We therefore impose the regularity conditions needed for our asymptotic analysis, namely cash-invariance, positive homogeneity, and $L^p$-Lipschitz continuity. While these assumptions do not imply coherence, coherent risk measures provide a natural and important class of examples: every finite coherent risk measure on $L^p$ is also $L^p$-Lipschitz continuous.

\pagebreak[2]
\begin{lemma}[{Finiteness and Lipschitz continuity of risk measures \cite[Corollary~2.6]{KainaRueschendorf2009}}]
\label{lem:::FiniteCoherentRiskMeasures}
If ${\rho \colon L^p\to (-\infty,\infty]}$ is a coherent risk measure for ${p \in [1, \infty)}$, then\/ $\rho$ is Lipschitz continuous (see \cref{asm:::LpLipschitz} below) if and only if it is finite.
\end{lemma}

\pagebreak[2]
The $L^p$-Lipschitz condition is particularly convenient for our purposes because it allows us to control differences in risk measures by the $L^p$-distance between the corresponding random variables. The assumption is therefore imposed not as a consequence of coherence, but as the regularity condition needed for the subsequent asymptotic analysis.

\pagebreak[2]
\begin{assumption}[$L^p$-Lipschitz continuity]
\label{asm:::LpLipschitz}

Assume that the risk measure $\rho$ is Lipschitz continuous on $L^p$, i.e., there exists some $C>0$ such that
\begin{equation}
\label{eq:::LpLipschitz}
|\rho(X)-\rho(Y)|
\le
C\|X-Y\|_{p},
\qquad X,Y\in L^p.
\end{equation}
\end{assumption}

\pagebreak[2]
The class of finite law-invariant coherent risk measures on $L^p$ --- which, by \cref{lem:::FiniteCoherentRiskMeasures}, are also $L^p$-Lipschitz continuous --- includes important examples such as Tail Value at Risk (TVaR), which is finite
on~$L^1$, since 
\begin{equation*}
  |\mathrm{TVaR}_\alpha(X)| \le \frac{1}{1-\alpha}\int_{[0,1]} |F^\leftarrow(u)|\,\mathrm{d}u =\frac{1}{1-\alpha}\|X\|_{1}.
\end{equation*}
Value at Risk (VaR), although widely used in practice, does not belong to this class and will therefore be treated separately.

\pagebreak[2]
\begin{definition}[{Value at Risk and Tail Value at Risk
}]\label{def:::VaR}
Let $X$ be a real-valued random variable and let $\alpha \in (0,1)$. The \emph{Value at Risk (VaR)} of $X$ at level $\alpha$ is defined by
\begin{equation}\label{eq:def VaR}
\VaR_\alpha(X) \coloneqq F^\leftarrow(\alpha) \coloneqq 
\inf \bigl\{\, x \in \mathbb{R} \;|\; 
\mathbb{P}[X \le x] \ge \alpha \,\bigr\}.
\end{equation}
If $X$ is integrable, the \emph{Tail Value at Risk (TVaR)} of $X$ at level $\alpha$ is given by
\begin{equation*}
\TVaR_\alpha(X) \coloneqq 
\frac{1}{1-\alpha}\int_{\alpha}^1 
\VaR_u(X) \,\mathrm{d}u.
\end{equation*}
\end{definition}

\pagebreak[2]
Since $\operatorname{TVaR}_\alpha(\Phi)$ admits the explicit expression $\operatorname{TVaR}_\alpha(\Phi) = \tfrac{\phi(\Phi^{-1}(\alpha))}{1-\alpha}$, the leading-order coefficient in the asymptotic cancellation cost is explicit for TVaR.
Other finite law-invariant coherent risk measures on $L^1$ include expectiles. As for finite coherent risk measures bounded on $L^p$ for some $p \in (1,\infty)$, we mention distortion Haezendonck--Goovaerts risk measures under some mild regularity assumption.

\pagebreak[2]
\begin{example}[{Expectiles, cf.\ \cite{BeDi17} and \cite{Whitney1987}}]
The expectile $\mu_\alpha$ for $\alpha \in (0,1)$ is uniquely characterised by the condition
\begin{equation*}
\alpha \mathbb{E}[(X - \mu_\alpha)^+] = (1 - \alpha) \mathbb{E}[(X - \mu_\alpha)^-].
\end{equation*}
An alternative representation of expectiles is given in \cite[Equation (2.7)]{Whitney1987} by
\begin{equation*}
\mu_\alpha(X) = \mathbb{E}[X] + \frac{2\alpha - 1}{1 - \alpha} \mathbb{E}[(X - \mu_\alpha)^+], \quad \alpha \in (0,1),
\end{equation*}
which follows directly from the identity $\mathbb{E}[(X - \mu_\alpha)^+] - \mathbb{E}[(X - \mu_\alpha)^-] = \mathbb{E}[X] - \mu_\alpha$.
See \cite[Proposition 2.1]{BeDi17} and \cite[Theorem 1]{Whitney1987} for the following properties of expectiles:

\pagebreak[2]
\begin{enumerate}[label=(\alph*)]
\item The expectiles $\mu_\alpha$ are strictly increasing in $\alpha$ on $(0,1)$.
\item They satisfy the symmetry relation $\mu_\alpha(-X) = -\mu_{1-\alpha}(X)$ for all $\alpha \in (0,1)$.
\item For all $\lambda > 0$ and $c \in \mathbb{R}$, the scaling and translation property
$\mu_\alpha(\lambda X + c) = \lambda \mu_\alpha(X) + c$ holds for all $\alpha \in (0,1)$.
\end{enumerate}
From these properties, we obtain that $\mu_\alpha(X - \mathbb{E}[X]) \ge 0$ for all $\alpha \in [1/2,1)$. Consequently, we derive the $L^1$-bound
\begin{equation*}
\mu_\alpha(X) = \mathbb{E}[X] + \mu_\alpha(X - \mathbb{E}[X])
\le \mathbb{E}[X] + \frac{2\alpha - 1}{1 - \alpha} \mathbb{E}[X^+]
\le \frac{\alpha}{1 - \alpha} \|X\|_{1}, \quad \alpha \in [1/2,1).
\end{equation*}
Finally, using the symmetry property, we conclude that $\mu_\alpha$ defines a coherent risk measure (see \cref{def:::RiskMeasure}) for all $\alpha \in [1/2,1)$ (cf. \cite[Introduction]{BeDi17}). Hence, expectiles for $\alpha \in [1/2,1)$ are finite law-invariant coherent risk measures.
\end{example}

\pagebreak[2]
\begin{example}[Distortion Haezendonck--Goovaerts risk measures \cite{goulard2025}]

For $\alpha \in [0,1)$ and $p \ge 1$, as well as a distortion function $g \colon [0,1] \to [0,1]$ (that is, $g$ is increasing, right-continuous, and satisfies $g(1)=1$), we formally define the \textit{Orlicz–Lorentz premium} of a non-negative random variable $X$ by
\begin{equation*}
\pi_{g,p,\alpha}(X)
\coloneqq
\biggl(\frac{1}{1-\alpha}
\int_{[0,1]} \bigl(F_X^{\leftarrow}(1-u)\bigr)^{p}\, g(\mathrm{d}u)
\biggr)^{1/p}.
\end{equation*}
Based on this premium, the \textit{distortion Haezendonck--Goovaerts risk measure} is formally defined as
\begin{equation*}
\rho_{g,p,\alpha}(X)
\coloneqq
\inf_{x \in \mathbb{R}}
\bigl( \pi_{g,p,\alpha}[(X-x)^+] + x \bigr).
\end{equation*}
Under the following regularity assumptions on $g$, the distortion Haezendonck–Goovaerts risk measure is well defined on some $L^p$ space and satisfies Assumption \ref{asm:::LpLipschitz}, as shown below.

\pagebreak[2]
\begin{itemize}
\item\textit{Domain:} Assume that $g$ is absolutely continuous and that the derivative $g'$ is bounded in $L^r(\mathrm{d}\lambda)$ for some $r \in [1,\infty]$, where $\lambda$ denotes the Lebesgue measure.\footnote{Note that for an absolutely continuous distortion function $g$, the derivative $g'$ is automatically bounded in $L^1(\mathrm{d}\lambda)$. However, this case is not particularly relevant for our purposes: when $r=1$, we have $r'=\infty$, and the resulting risk measure is only guaranteed to be bounded in $L^\infty$. Our asymptotic results, however, require boundedness in $L^p$ for some $p\in[1,\infty)$. Thus, the case $r=1$ falls outside the scope of our results. An example of a risk measure that is bounded in $L^\infty$ but not in any $L^p$ for $p<\infty$ is the essential supremum (see \cref{exm:::EssentialSupremum}).} Then, by H\"older's inequality, for any non-negative random variable~$X$
\begin{multline*}
\bigg( \int_{[0,1]} |F_X^{\leftarrow}(1-u)|^p \, g(\mathrm{d}u)\bigg)^{1/p} \\
=
\bigg(  \int_{[0,1]} |F_X^{\leftarrow}(1-u)|^p g'(u) \,\lambda(\mathrm{d}u)\bigg)^{1/p} \le
\| g' \|_{r}^{1/p} \| X \|_{{pr'}},
\end{multline*}
where $r'$ is the Hölder conjugate of $r$, i.e.\ $1/r+1/{r'}=1$. Therefore, the distortion Haezendonck--Goovaerts risk measure is well-defined on $L^{pr'}$.

\pagebreak[2]
\item\textit{$L^{pr'}$-boundedness:} Under the assumption that the distortion function $g$ is absolutely continuous with $L^r(\mathrm{d}\lambda)$-bounded derivative $g'$, note that 
\begin{equation*}
\rho_{g,p,\alpha}(X) \le \pi_{g,p,\alpha}[X^+] 
\le 
\biggl(\frac{\| g'\|_{r}}{1-\alpha}\biggr)^{1/p} \| X \|_{{pr'}} .
\end{equation*}
Since it is coherent (see \cite{goulard2025}), we can conclude that the distortion Haezendonck--Goovaerts risk measure is a finite law-invariant coherent risk measure on $L^{pr'}$.
\end{itemize}
In particular, choosing $g=\mathrm{id}$ and $p=1$ yields TVaR, while choosing $g=\mathrm{id}$ and arbitrary $p\ge1$ yields the classical Haezendonck–Goovaerts risk measure.
\end{example}

\section{Constant cancellation regime}
\label{sec:::ConstantCancellationRegime}

\subsection{The Cram\'er condition}

We now turn to the asymptotic analysis of the constant cancellation regime, where the number of cancelled policies~$m$ remains fixed as the portfolio size~$n$ tends to infinity. In the Gaussian benchmark, we have seen that the change in individual risk contribution is of order $n^{-3/2}$. We show that this rate remains valid for more general loss distributions under suitable regularity assumptions. In particular, the following condition will be imposed.

\pagebreak[2]
\begin{assumption}[Cram\'er's condition]
\label{ass:CramerCondition}

Let $X$ be an $\mathbb{R}$-valued random variable with characteristic function $\varphi(y)=\mathbb{E}[\exp(\mathrm{i} y X)]$ for $y \in \mathbb{R}$.
We say that $X$ satisfies Cram\'er's condition if $\limsup_{|y|\to\infty}|\varphi(y)|<1$.

\end{assumption}

Note that every random variable with an absolutely continuous component satisfies Cram\'er's condition. This is well known, and follows
easily from the Riemann-Lebesgue lemma.
For some results concerning singular continuous
distributions, see \cite{HuLa02,HuLa19,Ya13} and the references therein.
Purely atomic distributions never satisfy the Cram\'er condition.
This is also well known \cite[p.~207]{BhRa76}, and follows from the fact that
a characteristic function satisfying the condition cannot be almost-periodic
(Definition~VI.5.1 in~\cite{Ka04}),
while trigonometric polynomials and their uniform limits \emph{are}
almost-periodic (see p.~174 in~\cite{Ka04}).
In practice, life insurance contracts usually involve a payoff triggered
by the death of the customer, which can occur at any day. The common
actuarial practice of shifting this payment to the end
of the month or year is just an approximation. Thus, assuming
 a continuous component is not much of a restriction. If desired, results
on cancellation cost for distributions not satisfying the
Cram\'er condition could be derived
from  \cite{Bo18,DoFe23,Fel2,GovZ06,KoMc90}, and we do not expect essentially
different first-order asymptotics, but we leave this for
future research.

\subsection{Value at Risk}

Asymptotic expansions of $\Delta_{m,n}(X,\VaR)$ for~$m$ fixed follow from the Cornish--Fisher
expansion, which is the inversion of the  classical Edgeworth expansion. We state only the expansion based on the fourth moment, which suffices for first-order asymptotics of~$\Delta_{m,n}$. Higher-order expansions can be obtained under stronger moment assumptions.

\pagebreak[2]
\begin{lemma}[Classical Edgeworth expansion]
\label{cor:::ClassicalEdgeworthExpansion}
Let\/ $F$ be a standardised cdf with finite fourth moment satisfying Cram\'er's condition (see \cref{ass:CramerCondition}). Define\/ $F_n(x) = F^{*n}(\sqrt{n}\,x)$, where\/ $F^{*n}$ denotes the\/ $n$-fold convolution of\/ $F$. Then, for every\/ $x\in\mathbb{R}$,
\begin{equation}
\label{eq:::ClassicalEdgeworthExpansion}
F_n(x)=
\Phi(x)+\phi(x)
\biggl[
\frac{R_{1/2}(x)}{n^{1/2}}+\frac{R_1(x)}{n}+o\Bigl(\frac{1}{n}\Bigr)
\biggr],
\end{equation}
where
\begin{equation*}
R_{1/2}(x)
=
-\frac{\kappa_3}{6}H_2(x)
\quad\text{and}\quad
R_1(x)
=
-\Bigl(\frac{\kappa_4}{24}H_3(x)
+
\frac{\kappa_3^2}{72}H_5(x)
\Bigr)
\end{equation*}
with\/ $H_k$ denoting the\/ $k$-th Hermite polynomial, defined by\/ $H_k(x)=(-1)^k\phi^{(k)}(x)/\phi(x)$, and\/ $\kappa_k$ denoting the\/ $k$-th cumulant of\/ $F$, defined by\/ $\kappa_k = (-\mathrm{i})^{k} \frac{\partial^k}{\partial y^k}\log \varphi(y) |_{y=0}$ with\/ $\varphi(y) = \int \exp(\mathrm{i}yx)\,F(\mathrm{d}x)$.
\end{lemma}

\pagebreak[2]
\begin{proof}

A proof can be found in \cite[Chapter XVI, Section 4, Theorem 3]{Fel2} or \cite[Theorem 2.2]{Hal92}.
The expansion also follows from the more general \cref{thm:GeneralizedEdgeworthExpansion} below by taking $s=4$, $s'=0$, and $f=\mathbf{1}_{(-\infty,x]}$; the key step is the corresponding bound on~$\bar w_f$.\qedhere
\end{proof}

\pagebreak[2]
Inverting the Edgeworth expansion for~$F_n$ yields
\begin{equation}
\label{eq:::CornishFisher}
F_n^\leftarrow(\alpha) = \Phi^{-1}(\alpha) + \frac{Q_{1/2}(\Phi^{-1}(\alpha))}{n^{1/2}} + \frac{Q_{1}(\Phi^{-1}(\alpha))}{n} + o\Bigl(\frac{1}{n}\Bigr)
\end{equation}
for $\alpha \in (0,1)$ as $n \to \infty$, with
\begin{equation*}
Q_{1/2}(u) = \frac{\kappa_3}{6} (u^2-1)
\quad\text{and}\quad
Q_{1}(u) = \frac{\kappa_4}{24} (u^3 -3u) - \frac{\kappa_3^2}{36}(2u^3-5u).
\end{equation*}
This is the Cornish--Fisher expansion (see \cite[Theorem 2.4]{Hal92}). It allows us to directly derive the asymptotic behaviour of the change
in individual risk contribution for VaR.

\pagebreak[2]
\begin{theorem}[Asymptotics for constant cancellation under Value at Risk]
\label{thm:::AsymptoticsConstantVaR}

Let\/ $X$ be a random variable with standard deviation\/ $\sigma$ with finite fourth moment (i.e.\ \/$\mathbb{E}[|X|^4]<\infty$) satisfying Cram\'er's condition (see \cref{ass:CramerCondition}). Then, for $\alpha\in(0,1)$,
\begin{equation}
\Delta_{m,n}(X,\VaR_\alpha) = \frac{m \sigma }{2n^{3/2}}\Phi^{-1}(\alpha)  + o\biggl(\frac{1}{n^{3/2}}\biggr), \quad
m\geq 1,\ n\to\infty.
\end{equation}

\end{theorem}

\pagebreak[2]
\begin{proof}
For fixed $m \in \mathbb{N}$, by the Cornish--Fisher expansion \eqref{eq:::CornishFisher},
\begin{align*}
\Delta_{m,n}(X,\VaR_\alpha) &= \frac{\VaR_\alpha\big((n-m)^{1/2}\bar{X}_{n-m}\big)}{(n-m)^{1/2}}
      - \frac{\VaR_\alpha(n^{1/2}\bar{X}_{n})}{n^{1/2}} \notag \\
    &= \biggl(\frac{1}{(n-m)^{1/2}}-\frac{1}{n^{1/2}} \biggr) \Phi^{-1}(\alpha) + 
       \biggl(\frac{1}{n-m}-\frac{1}{n} \biggr) Q_{1/2}\bigl(\Phi^{-1}(\alpha)\bigr) \\&\quad+ 
       \biggl(\frac{1}{(n-m)^{3/2}}-\frac{1}{n^{3/2}} \biggr) Q_{1}\bigl(\Phi^{-1}(\alpha)\bigr) + o\biggl(\frac{1}{n^{3/2}}\biggr) \\
    & = \frac{ m}{2 n^{3/2}} \Phi^{-1}(\alpha) +o\biggl(\frac{1}{n^{3/2}}\biggr).
    \qedhere
\end{align*}
\end{proof}

\subsection{Law-invariant $L^1$-Lipschitz risk measures}

Unlike Value at Risk, general law-invariant risk measures cannot be represented in terms of a single quantile. Consequently, the Cornish--Fisher expansion is no longer sufficient for deriving asymptotic expansions of the cancellation cost. Instead, we require an approximation that allows us to evaluate expectations of sufficiently regular functions under the distribution of the standardised portfolio loss. To this end, we employ a generalized Edgeworth expansion due to Bhattacharya and Rao (see \cite[Theorem 20.1]{BhRa76}). Combined with the dual representation of the Wasserstein distance (i.e.\ Kantorovich--Rubinstein duality), this enables us to derive asymptotic expansions for law-invariant risk measures that are Lipschitz continuous on $L^1$. We first introduce the notation needed to state the generalized Edgeworth expansion.

\pagebreak[2]
\begin{definition}[Power bound and mean oscillation for functions]
For a real-valued function $f\colon \mathbb{R} \to \mathbb{R}$ and $s \in \mathbb{R}_+$ define
\begin{equation*}
M_{s}(f) \coloneqq
\left\{
\begin{array}{ll}
\sup_{x \in \mathbb{R}} (1+|x|^s)^{-1} |f(x)|, & s>0, \\
\sup_{x,y \in \mathbb{R}} |f(x) - f(y)|, & s=0.
\end{array}
\right .
\end{equation*}
Moreover, for a function $f \colon \mathbb{R} \to \mathbb{R}$ and $\varepsilon > 0$, we define its oscillation at $x$ and its mean oscillation with respect to a measure $\mu$ by

\begin{equation*}
w_f(x,\varepsilon)\coloneqq \sup_{y,z \in B_\varepsilon(x)} |f(z) - f(y)|
\quad\text{and}\quad
\bar w_f(\varepsilon,\mu) := \int w_f(x\,|\, \varepsilon) \; \mu(\mathrm{d}x),
\end{equation*}
where $B_\varepsilon(x) = \{ y \in \mathbb{R} \mid |y-x| < \varepsilon \}$.
\end{definition}
In the generalized Edgeworth expansion, we approximate the distribution of $F_n$ by a suitably constructed signed measure. In contrast to the central limit theorem (CLT) --- the leading Gaussian approximation --- the Edgeworth expansion yields higher-order corrections and thus explicit asymptotics in terms of the cumulants of the underlying distribution. This refinement is crucial for our derivation (in the constant cancellation regime). The higher-order correction terms in the generalized Edgeworth expansion are represented by signed measures obtained from derivatives of the Gaussian density. We briefly recall their construction in the one-dimensional case, which is sufficient for our purposes; the multidimensional case is given in \cite{BhRa76}.

\pagebreak[2]
\begin{definition}[Approximating signed measure]
\label{def:::ApproximatingSignedMeasure}
Assume we are given $\kappa =(\kappa_k)_{k \in \mathbb{N}} \in \mathbb{R}^\mathbb{N}$. Define the polynomials $\tilde P_s(z \mid \kappa )$ by the following identity between two formal power series
\begin{equation*}
1 + \sum_{s=1}^\infty \tilde P_s(z \mid \kappa ) u^s = 
\exp\Bigl( 
\sum_{s=1}^\infty \frac{\kappa_{s+2} z^{s+2}}{(s+2)!} u^s 
\Bigr) =
1+ \sum_{m=1}^\infty \frac{1}{m!}\Bigl(
\sum_{s=1}^\infty \frac{\kappa_{s+2} z^{s+2}}{(s+2)!} u^s
\Bigr)^m .
\end{equation*}
We obtain $P_s(-\phi \mid \kappa )$ as the Fourier transform of the function $\tilde P_s(\mathrm{i}t \mid \kappa ) \exp(-\tfrac{1}{2} t^2)$. This is given by formally substituting $(-1)^j \phi^{(j)}$ for $(\mathrm{i}t)^j$ in the polynomials $\tilde P_s(\mathrm{i}t \mid \kappa )$. Then $P_s(-\Phi \mid \kappa )$ is the signed measure on $\mathbb{R}$ whose density is $P_s(-\phi \mid \kappa )$.
\end{definition}

\pagebreak[2]
 \cref{def:::ApproximatingSignedMeasure} yields an explicit representation in terms of Hermite polynomials. We can calculate $\tilde P_s(z \mid \kappa)$ for $s \in \{1,2\}$ explicitly and conclude
\begin{equation*}
\tilde P_1(z \mid \kappa) = \frac{\kappa_3 z^3}{3!} 
\quad \text{and} \quad
\tilde P_2(z \mid \kappa)  = \frac{\kappa_4 z^4}{4!} + \frac{\kappa_3^2 z^6}{2!(3!)^2} .
\end{equation*}
Next we use the probabilistic Hermite polynomials defined by
\begin{equation}
\label{eq:::DefinitionOfProbHermitePolynomials}
H_n(x) \coloneqq (-1)^n \frac{\phi^{(n)}(x)}{\phi(x)}, \quad n \in \mathbb{N},
\end{equation}
where $\phi$ is the pdf of the standard normal distribution, to get the explicit form of $P_s(-\phi \mid \kappa)$ for $s \in \{0,1,2\}$. By the above construction, we can conclude that
\begin{multline*}
P_0( - \phi \mid \kappa )[x] = \phi(x), \quad
P_1(- \phi \mid \kappa )[x] = \phi(x) \frac{\kappa_3}{3!} H_3(x), \\ \text{and}\quad
P_2(- \phi \mid \kappa )[x] = 
\phi(x) \Bigl( \frac{\kappa_4}{4!} H_4(x) + \frac{\kappa_3^2}{2!(3!)^2} H_6(x) \Bigr) .
\end{multline*}
The first probabilistic Hermite polynomials are given by
\begin{equation*}
H_0(x) = 1,\quad
H_1(x) =x,\quad
H_2(x)= x^2-1,\quad
H_3(x)= x^3 - 3x,\quad
H_4(x)=x^4-6x^2+3
\end{equation*}
and
\begin{equation*}
H_5(x) = x^5 - 10x^3+15x, \quad
H_6(x) = x^6 - 15x^4 +45x^2 - 15 . 
\end{equation*}
With these notions in place, we can now recall the Edgeworth expansion in the form established by Bhattacharya and Rao in \cite[Theorem 20.1]{BhRa76}. Although their result is stated in the general finite-dimensional setting, we only consider the one-dimensional case here for simplicity. For many further references on the Edgeworth expansion, we refer to the introduction of~\cite{Ha25}.

\pagebreak[2]
\begin{theorem}[{Generalized Edgeworth expansion \cite[Theorem 20.1]{BhRa76}}]
\label{thm:GeneralizedEdgeworthExpansion}

Let\/ $(X_n)_{n\in\mathbb{N}}$ be an i.i.d.\ sequence of standardised\/ real-valued random variables with common distribution function~$F$, where:
\begin{itemize}
\item The distribution\/ $F$ satisfies Cram\'er's condition (see \cref{ass:CramerCondition}).
\item $F$ has finite\/ $s$-th absolute moment for some\/ $s\ge 3$ (i.e.\ $\mathbb{E}[|X_1|^s]<\infty$).
\end{itemize}
Denote by\/ $\kappa_k$ the\/ $k$-th cumulant for\/ $3 \le k \le s$.
Then  for every Borel-measurable function\/~$f \colon \mathbb{R} \to \mathbb{R}$ with\/ $M_{s'}(f) < \infty$ for some\/ $s' \in [\,0,s]$, we have
\begin{equation}
\label{eq:::ExtensionOfEdgeworthExtension}
\biggl| \int f(x) \;\mathrm{d}\Bigl[ F_n(x) - \sum_{r=0}^{s-2} n^{-r/2} P_r(-\Phi \mid \kappa )\Bigr]\biggr|
\le M_{s'}(f)\,\delta(n) + C\,\bar w_f(\,2\exp(-\eta n) \mid \Phi),
\end{equation}
where\/~$\delta(n) = o(n^{-(s-2)/2})$ as\/ $n\to\infty$ and $\eta>0$. Further,\/ $C$ and\/ $\delta(\cdot)$ as well as $\eta$ do not depend on the function\/ $f$.
\end{theorem}

\pagebreak[2]
\cref{thm:GeneralizedEdgeworthExpansion} provides an approximation of expectations of sufficiently regular test functions by Gaussian expectation together with higher-order correction terms. To transfer this approximation to law-invariant risk measures that are Lipschitz continuous in~$L^1$, we require estimates in an appropriate metric. This is achieved by combining the generalized Edgeworth expansion with the dual representation of the Wasserstein distance from optimal transport theory.

\pagebreak[2]
\begin{theorem}[{Kantorovich--Rubinstein duality \cite[Theorem 1.14]{Villani2003}}]
\label{thm:::KantorovichRubinstein}
Let\/~$\mathcal{X}$ be a Polish space and let\/~$d$ be a lower semi-continuous metric on $\mathcal{X}$. Then for integrable probability measures\/~$P$ and\/~$Q$
\begin{equation}
\label{eq:::KantorovichRubinstein}
\inf_{C \in \Pi(P,Q)} \int_{\mathcal{X} \times \mathcal{X}} d(x,y) \; C(\mathrm{d}x,\mathrm{d}y) =
\sup_{\substack{f \in L^1(|P-Q|), \\\| f \|_{\mathrm{Lip}} \le 1}} \int_{\mathcal{X}} f \, \mathrm{d}(P-Q),
\end{equation}
where\/~$\|f\|_\mathrm{Lip} \coloneqq \sup_{x\neq y} |f(x)-f(y)|/d(x,y)$.
\end{theorem}

\pagebreak[2]
The Kantorovich--Rubinstein duality identifies the Wasserstein distance of order one as the supremum over Lipschitz continuous test functions. Since the generalized Edgeworth expansion provides an approximation for precisely such test functions, it can be directly transferred to estimates in the Wasserstein distance. This allows us to derive the asymptotic behaviour of the cancellation cost for law-invariant risk measures that are Lipschitz continuous with respect to the $L^1$-norm. In particular, this class includes law-invariant coherent risk measures bounded in~$L^1$.

\pagebreak[2]
\begin{theorem}[Asymptotics for constant cancellation under $L^1$-Lipschitz continuity] 
\label{thm:::AsymptoticsConstantL1}

Assume that~$\rho$ is a law-invariant cash-invariant and positive homogeneous risk measure that is Lipschitz continuous on\/ $L^1$ (see \cref{asm:::LpLipschitz}). Let\/ $X$ be a random variable with standard deviation\/ $\sigma$ and finite fourth moment satisfying Cram\'er's condition (see \cref{ass:CramerCondition}). Then
\begin{equation}
\label{eq:::ConstantL1}
\Delta_{m,n}(X,\rho) = 
\frac{m \sigma}{2} 
\frac{ \rho(\Phi)}{n^{3/2}}  + o\biggl(\frac{1}{n^{3/2}}\biggr), 
\quad m\geq 1,\ n\to\infty.
\end{equation}
\end{theorem}

\pagebreak[2]
\begin{proof}
Without loss of generality (see~\eqref{eq:normalize}), assume that $X$ is standardised. Fix $m \in \mathbb{N}$ and rewrite $\Delta_{m,n}$ as 
\begin{align}
\Delta_{m,n}(X,\rho) &=
\frac{1}{n^{1/2}} \big( \rho ( F_{n-m}) - \rho(F_n) \big) 
+
\bigg( \frac{1}{(n-m)^{1/2}}-\frac{1}{n^{1/2}}\bigg)
 \big(\rho(F_{n-m}) -\rho(\Phi) \big) \notag \\
 &\qquad+ \bigg( \frac{1}{(n-m)^{1/2}}-\frac{1}{n^{1/2}}\bigg)
 \rho(\Phi). \label{eq:::ProofAsympL1Decomp}
\end{align}
By $L^1$-Lipschitz continuity together with law-invariance, conclude
\begin{align}
\bigl|\rho ( F_{n-m}) - \rho( F_n) \bigr| 
&\le C
\inf_{Y_{n-m},Y_n \in \Pi(F_{n-m},F_n)} \| Y_{n-m}-Y_n\|_{1} \notag \\
&=
C\sup_{\|f\|_{\mathrm{Lip}}\le 1} 
\bigg(
\int f(x) \, F_{n-m}(\mathrm{d}x) - \int f(x) \, F_{n}(\mathrm{d}x)
\bigg), \label{eq:rho by sup}
\end{align}
where $C$ is the Lipschitz constant from \cref{asm:::LpLipschitz}, and the last equality follows from \cref{thm:::KantorovichRubinstein}. 
Note that
\begin{equation*}
\int (f(x)+c) \, F_{n-m}(\mathrm{d}x) - \int (f(x)+c) \, F_n(\mathrm{d}x)
= \int f(x) \, F_{n-m}(\mathrm{d}x) - \int f(x) \, F_n(\mathrm{d}x)
\end{equation*}
for all $c \in \mathbb{R}$, and so adding a constant to $f$ does not change the value of the expression. This allows us to impose the additional restriction $f(0)=0$ in the supremum. 
Note that the quantity $\delta(n)$ in \cref{thm:GeneralizedEdgeworthExpansion} is independent of the choice of $f$. Thus \cref{thm:GeneralizedEdgeworthExpansion} (for $s=4$) implies for all functions $f$ with $\|f\|_{\mathrm{Lip}}\le 1$ and $f(0)=0$ that
\begin{multline*}
\biggl|
\int f(x)\,\mathrm{d}
\biggl[
F_n(x) 
- \Phi(x) 
- \frac{\phi(x)}{n^{1/2}}\frac{\kappa_3}{6}H_2(x)
+\frac{\phi(x)}{n}
\Bigl( \frac{\kappa_4}{24}H_3(x) + \frac{\kappa_3^2}{72} H_5(x)\Bigr)
\biggr]
\biggr| \\
\le M_s(f) \delta(n) + C \bar w_f(2\exp(-\eta n) \mid \Phi).
\end{multline*}
For any function $f$ with $\|f\|_{\mathrm{Lip}}\le 1$ and $f(0)=0$, we conclude 
\begin{align*}
M_s(f) = \sup_{x \in \mathbb{R}}(1+|x|^s)^{-1}|f(x)| \le \sup_{x\in\mathbb{R}}(1+|x|^s)^{-1}|x| 
=
\frac{(s-1)^{(s-1)/s}}{s} < \infty
\end{align*}
and
\begin{equation*}
\bar w_f(\varepsilon \,|\, \Phi) =
\int_\mathbb{R} \sup_{y,z \in B_{\varepsilon}(x)} |f(y)-f(z)| \, \mathrm{d}\Phi(x) \le 2\varepsilon, \quad \varepsilon >0.
\end{equation*}
In particular, both expressions are bounded independently of the function $f$. Since $M_s(f)$ and $\bar w_f(\cdot,\Phi)$ are uniformly bounded over all admissible $f$, the remainder estimate in \cref{thm:GeneralizedEdgeworthExpansion} is uniform over this class. In particular this yields
\begin{equation}
\label{eq:::ApproximationEdgeworth}
\int_\mathbb{R} f(x) \,F_n(\mathrm{d}x) =
\int_\mathbb{R} f(x) \,\Phi(\mathrm{d}x) +
\frac{1}{n^{1/2}} I_{1/2}(f) +
\frac{1}{n} I_1(f) + \delta_f(n),
\end{equation}
with $\sup_{\|f\|_\mathrm{Lip}\le 1, f(0)=0} \delta_f(n) = o(n^{-1})$, where by using the definition of probabilistic Hermite polynomials
\begin{equation*}
I_{1/2}(f) = \int_\mathbb{R} f(x) \frac{\kappa_3 H_3(x)}{6}\,  \Phi(\mathrm{d}x)
\quad\text{and}\quad
I_1(f) = \int_\mathbb{R} f(x) 
\Bigl( 
\frac{\kappa_4 H_4(x)}{24}+\frac{\kappa_3^2H_6(x)}{72}
\Bigr)\,\Phi(\mathrm{d}x).
\end{equation*}
Insert~\eqref{eq:::ApproximationEdgeworth} into~\eqref{eq:rho by sup} to obtain
\begin{multline*}
|\rho(F_{n-m}) - \rho(F_n)| \le
C\biggl(\frac{1}{(n-m)^{1/2}}-\frac{1}{n^{1/2}}\biggr)
\sup_{\substack{\|f\|_\mathrm{Lip}\le 1 \\ f(0)=0}} |I_{1/2}(f)| \\
\quad + 
C
\biggl(\frac{1}{n-m} - \frac{1}{n}\biggr) 
\sup_{\substack{\|f\|_\mathrm{Lip}\le 1 \\ f(0)=0}} |I_{1}(f)| + o\Bigl(\frac{1}{n}\Bigr),
\end{multline*}
where the suprema are finite, since $|f(x)|\le |x|$ for all $f$ with $\|f\|_{\mathrm{Lip}}\le 1$ and $f(0)=0$ and thus
\begin{align*}
|I_{1/2}(f)| &\le \int |x| \frac{|\kappa_3 H_3(x)|}{6} \Phi(\mathrm{d}x) < \infty, \\
 |I_{1}(f)| &\le \int |x| \biggl(\frac{|\kappa_4 H_4(x)|}{24} + \frac{|\kappa_3^2 H_6(x)|}{72} \biggr) \Phi(\mathrm{d}x) < \infty.
\end{align*}
Furthermore,
\begin{equation}\label{eq:1/ expans}
\frac{1}{(n-m)^{1/2}}-\frac{1}{n^{1/2}} = \frac{m}{2n^{3/2}} + o\biggl( \frac{1}{n^{3/2}}\biggr) 
\quad\text{and}\quad
\frac{1}{n-m} - \frac{1}{n} = \frac{m}{n^2} + o\biggl(\frac{1}{n^2}\biggr),
\end{equation}
and therefore the first term on the right hand side of~\eqref{eq:::ProofAsympL1Decomp}
satisfies
\begin{equation}
\label{eq:::ProofAsympL1DecompFirstEstim}
\bigl|\rho ( F_{n-m}) - \rho( F_n) \bigr| = o\Bigl(\frac{1}{n}\Bigr).
\end{equation}
Law-invariance together with the Kantorovich–Rubinstein duality (see \cref{thm:::KantorovichRubinstein}) implies
\begin{align*}
|\rho(F_{n-m})- \rho(\Phi) | &\le \inf_{Y_{n-m},Z \in \Pi(F_{n-m},\Phi)} \| Y_{n-m}-Z\| \\
&= \sup_{\substack{\|f\|_{\mathrm{Lip}}\le 1, \\ f(0)=0}} \biggl( 
\int f(x) \, F_{n-m}(\mathrm{d}x) - \int f(x) \,\Phi(\mathrm{d}x)
\biggr). 
\end{align*}
Using again \eqref{eq:::ApproximationEdgeworth} allows to estimate
the second term on the right hand side of~\eqref{eq:::ProofAsympL1Decomp}:
\begin{multline}
\label{eq:::ProofAsympL1DecompSecondEstim}
|\rho(F_{n-m})- \rho(\Phi) | \le
\frac{1}{(n-m)^{1/2}} \sup_{\substack{\|f\|_{\mathrm{Lip}}\le 1,\\ f(0)=0}} |I_{1/2}(f)| \\
+
\frac{1}{n-m} \sup_{\substack{\|f\|_{\mathrm{Lip}}\le 1,\\ f(0)=0}} |I_{1}(f)| + o\biggl(\frac{1}{n}\biggr) 
=
O\biggl(\frac{1}{n^{1/2}}\biggr).
\end{multline}
Now apply the first
equality in~\eqref{eq:1/ expans} to the last term
of~\eqref{eq:::ProofAsympL1Decomp},  together with \cref{exm:::NormalDistribution}, to prove~\eqref{eq:::ConstantL1}.
\end{proof}

\pagebreak[2]
Note that finiteness of some moments, as assumed in \cref{thm:::AsymptoticsConstantL1} and subsequent results, is no severe restriction. Life insurers usually do not allow customers to choose arbitrarily large covers, and so the distribution of~$X$ may be assumed to have bounded support. Without these moment assumptions, $\Delta_{m,n}$ need not be of order $n^{-3/2}$, but the corresponding heavy-tailed loss distributions are
of little relevance for life insurance.
\begin{example}\label{ex:stable}
  Suppose that~$X$ is centred stable, with stability parameter $\hat{\alpha} \in (0,2)$
  and characteristic function $\varphi(y)=e^{-|y|^{\hat{\alpha}}} $. Then $\bar{X}_n$
  has the same distribution as $n^{1/\hat{\alpha} -1}X$. In particular,
  for $\hat{\alpha} =1$ we have $\bar{X}_n \overset{\mathrm{d}}{=}X$ for all~$n$, and so
  \[
    \Delta_{m,n}(X,\rho)=\rho(\bar{X}_{n-m}) - \rho(\bar{X}_n)=0, \quad m\geq1, n\geq2,
  \]
  for any law-invariant risk measure for which $\rho(X)$ is finite. For $1\neq\hat{\alpha}  \in (0,2)$ and $m\geq1$ fixed, we find
  \[
    \Delta_{m,n}(X,\rho) =
    \big((n-m)^{1/\hat{\alpha} -1}-n^{1/\hat{\alpha} -1} \big) \rho(X)
    \sim \frac{\hat{\alpha} -1}{\hat{\alpha} } m\rho(X) n^{1/\hat{\alpha} -2}.
  \]
  E.g., for $\rho=\mathrm{VaR}_{\alpha}$, the change in individual risk contribution
  is well-defined for all $\hat{\alpha} \in(0,2)$, and may even be negative.
\end{example}

\subsection{Extension to law-invariant $L^p$-Lipschitz risk measures for $p>1$}

The previous theorem relies on the Kantorovich--Rubinstein duality and therefore exploits the dual characterisation of the Wasserstein distance of order one. To extend the result to risk measures that are Lipschitz continuous on $L^p$ for $p\in(1,2)$, we combine Wasserstein estimates of different orders through an interpolation argument.

\pagebreak[2]
\begin{definition}[Wasserstein distance]
Let $(\mathcal{X} ,d)$ be a separable metric space and $p \in [\,1,\infty)$. Then 
\begin{equation}
\label{eq:::WassersteinMetric}
W_p(P,Q) = \bigg(\inf_{C \in \Pi(P,Q)} \int_{\mathcal{X} \times \mathcal{X}} d(x,y)^p \,C(\mathrm{d}x,\mathrm{d}y) \bigg)^{1/p}, \quad P,Q \in \mathcal{M}_1^p(\mathcal{X}),
\end{equation}
defines a metric on $\mathcal{M}_1^p(\mathcal{X}) \coloneqq \{ P \in \mathcal{M}_1(\mathcal{X}) \mid \int d(x_0,x)^p \,P(\mathrm{d}x) <\infty \text{ for some } x_0 \in \mathcal{X} \}$.
\end{definition}

\pagebreak[2]
The Wasserstein distance is indeed a metric on $\mathcal{M}_1^p(\mathcal{X})$ (see \cite[Theorem 7.3(a)]{Villani2003}). We refer to this metric as the Wasserstein metric of order~$p$. On $\mathbb{R}$, Wasserstein distances of different orders can be related through an interpolation argument.

\pagebreak[2]
\begin{lemma}[Interpolation of Wasserstein distances]
\label{lem:::WassersteinInterpolation}
For $1<p<r$ let $P,Q\in\mathcal{M}_1^r(\mathbb{R})$. Then
\begin{equation*}
W_p(P,Q)
\le
W_1(P,Q)^\alpha W_r(P,Q)^{1-\alpha},
\end{equation*}
where
\begin{equation*}
\alpha=\frac{r-p}{p(r-1)}.
\end{equation*}
\end{lemma}

\pagebreak[2]
\begin{proof}
Since $P$ and $Q$ are one-dimensional, the monotone (quantile) coupling is optimal for $W_s(P,Q)$ for every $s\ge1$ (see \cite[Remark 2.19(ii)]{Villani2003}. Let $C$ denote this coupling.
Then the estimate follows from an interpolation estimate for $L^p$-norms (see exercises in \cite[Chapter 8]{Wheeden2015}); for completeness, we give a proof. Consider $\frac{1}{p}=\alpha+\frac{1-\alpha}{r}$ and define
\begin{equation*}
q_1=\frac{1}{\alpha p}
\quad\text{and}\quad
q_2=\frac{r}{(1-\alpha)p}.
\end{equation*}
Then $\frac{1}{q_1}+\frac{1}{q_2}=p\left(\alpha+\frac{1-\alpha}{r}\right)=1$ and $q_1,q_2>1$. Hence, by Hölder's inequality,
\begin{align*}
W_p(P,Q) &= \Bigl(\int |x-y|^p\,C(\mathrm{d}x,\mathrm{d}y)\Bigr)^{1/p}
=
\Bigl(
\int |x-y|^{\alpha p}|x-y|^{(1-\alpha)p}
\,C(\mathrm{d}x,\mathrm{d}y)
\Bigr)^{1/p}\\
&\le
\Bigl(\int |x-y|\,C(\mathrm{d}x,\mathrm{d}y)\Bigr)^\alpha
\Bigl(\int |x-y|^r\,C(\mathrm{d}x,\mathrm{d}y)\Bigr)^{(1-\alpha)/r} \\
&= W_1(P,Q)^\alpha W_r(P,Q)^{1-\alpha}. \qedhere
\end{align*}
\end{proof}

\pagebreak[2]
The preceding results provide the key ingredient required to extend the constant cancellation asymptotics to risk measures that are Lipschitz continuous on $L^p$ for $p\in(1,2)$. The interpolation result in \cref{lem:::WassersteinInterpolation} allows us to combine the order-$1$ Wasserstein estimate, obtained from the generalized Edgeworth expansion, with a higher-order Wasserstein estimate based on the optimal rate in the central limit theorem for the Wasserstein distance (see \cite[Theorem~1.1]{Bobkov2018}). This yields the required bound for the Wasserstein distance of order $p$. Combining this estimate with the $L^p$-Lipschitz continuity of the risk measure then gives the following result. The restriction $p<2$ arises from the interpolation argument, which requires the interpolation exponent $\alpha$ to exceed $1/2$ in order to obtain the required rate. Since $\alpha<1/p$, this condition can only be satisfied for $p<2$.

\pagebreak[2]
\begin{theorem}[Asymptotics for constant cancellation under $L^p$-Lipschitz continuity]
\label{thm:::AsymptoticConstantLp}

Assume that~$\rho$ is a law-invariant, cash-invariant and positive homogeneous risk measure that is Lipschitz continuous on\/ $L^p$ (see \cref{asm:::LpLipschitz}) for some $p \in [1,2) $. 
Let~$X$ be a random variable with standard deviation\/ $\sigma$ and finite $k$-th absolute moment for some\/ $k > \tfrac{4-p}{2-p}$, satisfying Cram\'er's condition (see \cref{ass:CramerCondition}).
Then
\begin{equation}
\label{eq:::ConstantLp}
\Delta_{m,n}(X,\rho) = 
\frac{m}{2} 
\frac{\sigma \rho(\Phi)}{n^{3/2}}  + o\biggl( \frac{1}{n^{3/2}}\biggr), 
\quad m\geq 1,\ n\to\infty.
\end{equation}
\end{theorem}

\pagebreak[2]
\begin{proof}
We consider the case $p \in (1,2)$, since the case $p=1$ was treated in \cref{thm:::AsymptoticsConstantL1}. Without loss of generality, assume that $X$ is standardised. Fix $m \in \mathbb{N}$ and rewrite $\Delta_{m,n}$ as 
\begin{align}
\Delta_{m,n}(X,\rho) &=
\frac{1}{n^{1/2}} \big( \rho ( F_{n-m}) - \rho(F_n) \big) 
+
\bigg( \frac{1}{(n-m)^{1/2}}-\frac{1}{n^{1/2}}\bigg)
 \big(\rho(F_{n-m}) -\rho(\Phi) \big) \notag \\
 &\qquad+ \bigg( \frac{1}{(n-m)^{1/2}}-\frac{1}{n^{1/2}}\bigg)
 \rho(\Phi). \label{eq:::ProofAsympLpDecomp}
\end{align}
We first consider the first term on the right-hand side of \eqref{eq:::ProofAsympLpDecomp}. By \cref{lem:::WassersteinInterpolation} with $r \coloneqq k-2 > \tfrac{p}{2-p} > p$, we conclude
\begin{equation*}
W_p(F_n,F_{n-m}) \le W_1(F_n,F_{n-m})^\alpha W_r ( F_n, F_{n-m})^{1-\alpha},
\end{equation*}
where $\alpha = \tfrac{r-p}{p(r-1)} > \tfrac{1}{2}$. Similarly to \cref{thm:::AsymptoticsConstantL1} we can show, since $k > \frac{4-p}{2-p}>3$, that
\begin{equation*}
W_1(F_{n-m},F_n) = O\biggl(\frac{1}{n^{3/2}}\biggr). 
\end{equation*}
Since $X$ has finite absolute moments of order $r+2=k$, \cite[Theorem 1.1]{Bobkov2018} yields $W_r(F_n,\Phi) = O(n^{-1/2})$ and thus
\begin{equation*}
W_r(F_n,F_{n-m}) \le W_r(F_{n-m},\Phi) + W_r(\Phi, F_{n}) = O \biggl(\frac{1}{n^{1/2}}\biggr).
\end{equation*}
Combining these estimates gives
\begin{equation*}
W_p(F_n,F_{n-m}) = O\biggl(\frac{1}{n^{\alpha(3/2) + (1-\alpha)(1/2)}}\biggr) =
O\biggl(\frac{1}{n^{1/2 + \alpha}}\biggr) = o\biggl( \frac{1}{n}\biggr).
\end{equation*}
Hence by Lipschitz continuity
\begin{equation}
\label{eq:::EstimateFirstConstantLp}
|\rho(F_{n-m}) - \rho(F_n)|  = o\biggl(\frac{1}{n}\biggr).
\end{equation}
We now estimate the second term on the right-hand side of~\eqref{eq:::ProofAsympLpDecomp}. The CLT in Wasserstein distance (see \cref{lem:::CLTinWassersteinDistance} below, or \cite[Theorem 1.1]{Bobkov2018}, because the stronger moment assumptions are fulfilled) implies $W_p(F_{n-m},\Phi) \to 0$. Hence by Lipschitz continuity and~\eqref{eq:1/ expans},
\begin{equation}
\label{eq:::EstimateSecondConstantLp}
\biggl(\frac{1}{(n-m)^{1/2}}- \frac{1}{n^{1/2}}\biggr) \bigl( \rho(F_{n-m}) - \rho(\Phi)\bigr) = o\biggl(\frac{1}{n^{3/2}}\biggr).
\end{equation}
It remains to insert~\eqref{eq:::EstimateFirstConstantLp} and \eqref{eq:::EstimateSecondConstantLp} into \eqref{eq:::ProofAsympLpDecomp} to prove, together with \cref{exm:::NormalDistribution}, the statement~\eqref{eq:::ConstantLp}.
\end{proof}

\pagebreak[2]
\section{Proportional cancellation regime}
\label{sec:::ProportionalCancellationRegime}

\subsection{Value at Risk}

We next consider the proportional cancellation regime, where the number of cancelled policies grows proportionally with the size of the portfolio. More precisely, we assume that $m=\lfloor \theta n \rfloor$ for some fixed $\theta \in (0,1)$. In contrast to the constant cancellation regime, the two remaining portfolio sizes $n$ and $n-m$ differ on the same scale as the original portfolio. Consequently, the standardised distributions of the two portfolios converge to Gaussian limits with different scaling factors, and the cancellation cost already appears at the leading order of the central limit approximation.

Hence, unlike in the constant cancellation regime, higher-order corrections from the Edgeworth expansion are not required to identify the dominant asymptotic behaviour. Instead, the main contribution arises from the homogeneity property of the risk measure and the CLT scaling. We first establish the result for the VaR and then discuss extensions to the risk measures considered in the previous section.

\pagebreak[2]
\begin{lemma}[Asymptotics for proportional cancellation under Value at Risk]
\label{lem:::AsymptoticProportionalVaR}

Let $X$ be a random variable with standard deviation $\sigma$. Then
\begin{equation*}
\Delta_{\theta,n}(X;\VaR) = 
\frac{1-(1-\theta)^{1/2}}{(1-\theta)^{1/2}} \frac{\sigma\VaR(\Phi)}{n^{1/2}}
+ o\biggl( \frac{1}{n^{1/2}} \biggr), \quad \theta\in(0,1),\ n\to\infty.
\end{equation*}
\end{lemma}

\pagebreak[2]
\begin{proof}
Without loss of generality assume that $X$ is standardised and let $F_n$ denote the distribution function of the standardised portfolio loss. The CLT implies
$F_n \overset{\mathrm{d}}{\rightarrow} \Phi$,
where~$\Phi$ is the standard normal distribution function. 
Since $\Phi$ is continuous and strictly increasing, it follows that $\operatorname{VaR}_\alpha(F_n) \to \operatorname{VaR}_\alpha(\Phi)$.
Indeed, writing $q_\alpha = \Phi^{-1}(\alpha)$, weak convergence and the continuity of $\Phi$ imply that, for every $\varepsilon>0$ and sufficiently large $n \in \mathbb{N}$, $F_n(q_\alpha-\varepsilon) < \alpha < F_n (q_\alpha + \varepsilon)$, and thus $q_\alpha-\varepsilon\leq F_n^{\leftarrow}(\alpha)\leq q_\alpha+\varepsilon$. Hence
\begin{multline*}
\Delta_{\theta,n}(X,\VaR) = 
\Delta_{\theta,n}(\Phi,\VaR) \\
+\underbrace{\frac{1}{n^{1/2}}\bigl(
\VaR(\Phi)-\VaR(F_n)\bigr)
 + \frac{1}{(n-\lfloor\theta n\rfloor)^{1/2}}\bigl( \VaR(F_{n-\lfloor\theta n\rfloor}) - \VaR(\Phi)\bigr)}_{=o(n^{-1/2})},
\end{multline*}
which, together with \cref{exm:::NormalDistribution}, proves the claim.
\end{proof}

\pagebreak[2]
\subsection{Law-invariant $L^p$-continuous risk measures}
The previous result relies on the continuity of Value at Risk with respect to weak convergence at the Gaussian limit. For general law-invariant risk measures, however, weak convergence is insufficient to guarantee convergence of the risk measure. For instance, TVaR is not continuous with respect to weak convergence.\footnote{Consider the random variables $X_n = n Y_n$, where $Y_n \sim \mathrm{Bern}(1/n)$. Then $X_n \to 0$ weakly, while, for $1/n<1-\alpha$,  $\mathrm{TVaR}_\alpha(X_n) = \frac{1}{1-\alpha} \mathbb{E}[X_n] = \frac{1}{1-\alpha} \not\to 0 = \mathrm{TVaR_\alpha(0)}$.} We therefore require a stronger mode of convergence compatible with the continuity properties imposed on the risk measures considered in the previous section. In particular, convergence in the Wasserstein distance of order $p$ is sufficient for risk measures that are continuous on $(\mathcal{M}_1^p(\mathbb{R}),W_p)$. This motivates the use of a Wasserstein central limit theorem, which we state next.

\pagebreak[2]
\begin{lemma}[{Central Limit Theorem in Wasserstein distance, \cite[Theorem 3.2]{Johnson2005}}]
\label{lem:::CLTinWassersteinDistance}

Let\/ $p \in [1,\infty)$ and let\/ $(X_k)_{k \in \mathbb{N}}$ be an i.i.d.\ sequence of real-valued random variables with\/ $\mathbb{E}[X_1] = 0$, variance\/ $\sigma^2 \in (0,\infty)$, and finite\/ $p$-th moment, i.e.\ \/$\mathbb{E}[|X_1|^p] < \infty$. Denote by $F_n$ the distribution of the normalised partial sums $\frac{1}{\sigma n^{1/2}} \sum_{k=1}^n X_k$. Then
\begin{equation}
W_p(F_n,\Phi) \longrightarrow 0, \qquad \text{as } n \to \infty.
\end{equation}
\end{lemma}

\pagebreak[2]
\begin{proof}
For $p \ge 2$, the result is given in \cite[Theorem 3.2]{Johnson2005}. For $p \in [1,2)$ apply the result to the case $p=2$ and use the ordering of Wasserstein distances $W_p \le W_q$ for all $p \le q$ in $[1,\infty)$ (see \cite[Section 7.1.2]{Villani2003}).
\end{proof}

The rate of convergence in this central limit theorem has also been studied in the literature. Under the assumptions of \cref{lem:::CLTinWassersteinDistance}, \cite[Introduction]{Rio2009} shows that the rate $n^{1/p-1/2}$ is optimal in general. In particular, attaining the optimal rate $O(n^{-1/2})$ requires stronger moment assumptions. For this, we rely on \cite[Theorem 1]{Bobkov2018}, which establishes this rate under the existence of a finite $(p+2)$-th moment.

\pagebreak[2]
\begin{theorem}[Asymptotics for proportional cancellation under $W_p$-continuity]
\label{thm:::AsymptoticProportionalLp}

Let~$\rho$ be a law-invariant cash-invariant and positive homogeneous risk measure  that is continuous on\/ $(\mathcal{M}_1^p(\mathbb{R}),W_p)$ for some\/ $p\in [1,\infty)$. 
Let\/ $X$ be a random variable with standard deviation\/ $\sigma$ and finite $p$-th moment. Then
\begin{equation*}
\Delta_{\theta,n}(X,\rho) = 
\frac{1-(1-\theta)^{1/2}}{(1-\theta)^{1/2}} 
\frac{\sigma \rho(\Phi)}{n^{1/2}}  + o\biggl( \frac{1}{n^{1/2}}\biggr),  \quad \theta\in(0,1),\ n\to\infty.
\end{equation*}
\end{theorem}

\pagebreak[2]
The class of risk measures considered in the theorem includes, in particular, all law-invariant finite coherent risk measures discussed in the preliminaries (see \cref{lem:::FiniteCoherentRiskMeasures}). Note, however, that subadditivity is not used in the proof. We therefore state the theorem in the more general form above.

\pagebreak[2]
\begin{proof}[Proof of \cref{thm:::AsymptoticProportionalLp}]
Without loss of generality assume that $X$ is standardised. Using positive homogeneity, we obtain
\begin{equation*}
\Delta_{\theta,n}(X,\rho) =
\Delta_{\theta,n}(\Phi,\rho) +
\frac{1}{n^{1/2}} \bigl(\rho(\Phi)-\rho(F_n)\bigr) +
\frac{1}{(n-\lfloor{\theta n}\rfloor)^{1/2}}
\bigl( \rho(F_{n-\lfloor\theta n\rfloor}) - \rho(\Phi) \bigr).
\end{equation*}
By \cref{lem:::CLTinWassersteinDistance}, $W_p(F_n,\Phi)\to0$ and $W_p(F_{n-\lfloor\theta n\rfloor},\Phi)\to0$.
Since $\rho$ is continuous with respect to the Wasserstein distance of order $p$, it follows that
\begin{equation*}
\rho(F_n)\to\rho(\Phi)
\quad\text{and}\quad
\rho(F_{n-\lfloor\theta n\rfloor})\to\rho(\Phi).
\end{equation*}
Consequently, $n^{-1/2}(\rho(\Phi)-\rho(F_n)) =o(n^{-1/2})$, and similarly for the second remainder term. Together with the explicit expression for $\Delta_{\theta,n}(\Phi,\rho)$ established in \cref{exm:::NormalDistribution}, this proves the theorem.
\end{proof}

\pagebreak[2]
The previous theorem establishes the first-order asymptotic behaviour of the proportional cancellation regime under Wasserstein continuity. Under the stronger moment assumptions of \cite[Theorem 1.1]{Bobkov2018}, the Wasserstein central limit theorem provides the sharper convergence rate discussed above. This allows us to slightly improve the remainder term in \cref{thm:::AsymptoticProportionalLp} under the additional assumption of Lipschitz continuity of the risk measure (e.g.\ finite law-invariant coherent risk measures on $L^p$ by \cref{lem:::FiniteCoherentRiskMeasures}).
Let $\rho$ be a law-invariant, cash-invariant, and positively homogeneous risk measure that is Lipschitz continuous on $L^p$ (see \cref{asm:::LpLipschitz}) for some $p\in[1,\infty)$, and let $X$ be a random variable with standard deviation $\sigma$ and finite $(p+2)$-th moment. Without loss of generality, assume that $X$ is standardised. By \cite[Theorem 1.1]{Bobkov2018}, $W_p(F_n,\Phi)=O(n^{-1/2})$. Together with law invariance and Lipschitz continuity, this gives
\begin{equation*}
\frac{1}{n^{1/2}}\bigl(\rho(\Phi)-\rho(F_n)\bigr)
=O(n^{-1})
\end{equation*}
and
\begin{equation*}
\frac{1}{(n-\lfloor\theta n\rfloor)^{1/2}}
\bigl(\rho(F_{n-\lfloor\theta n\rfloor})-\rho(\Phi)\bigr)
=O\bigl((n-\lfloor\theta n\rfloor)^{-1}\bigr).
\end{equation*}
Since $(n-\lfloor\theta n\rfloor)^{-1}=O(n^{-1})$ for fixed $\theta\in(0,1)$, we obtain, together with \cref{exm:::NormalDistribution},
\[
\Delta_{\theta,n}(X,\rho)
=
\frac{1-(1-\theta)^{1/2}}{(1-\theta)^{1/2}}
\frac{\sigma\rho(\Phi)}{n^{1/2}}
+
O(n^{-1}).
\]
Thus, under the stronger moment and Lipschitz assumptions, the $o(n^{-1/2})$ remainder in \cref{thm:::AsymptoticProportionalLp} can be sharpened to $O(n^{-1})$.

\pagebreak[2]
\subsection{Proportional cancellation in a heterogeneous portfolio}
\label{subsec:::ProportionalCancellationInheterogeneousPortfolio}

The preceding analysis considers a homogeneous portfolio in which the individual risks are identically distributed. We now extend the analysis to a heterogeneous portfolio consisting of a finite number of independent segments. Within each segment, the individual risks are assumed to be identically distributed, while the distributions may differ across segments. In addition, the proportion of cancelled contracts may vary between segments.
We index the segments by $j\in\{1,\ldots,l\}$, where $l\in\mathbb{N}$, and denote by $w_j$ the proportion of the total portfolio represented by segment $j$, with
\begin{equation*}
w_j > 0
\quad \text{and} \quad
\sum_{j=1}^{l}w_j=1.
\end{equation*}
Let $n\in\mathbb{N}$ denote the total portfolio size and let $n_j\in\mathbb{N}$ denote the number of risks in segment $j$, where
\begin{equation*}
n_j=w_jn+o(n),\qquad n\to\infty,
\end{equation*}
with $\sum_{j=1}^l n_j=n$. The $o(n)$ term allows for deviations from the limiting segment proportions, including those arising from rounding to integer segment sizes. For each segment, let $\sigma_j^2$ denote the variance of an individual risk and let $\theta_j\in [0,1)$ denote the proportion of risks cancelled in that segment. We assume that the random variables $X_{j,k}$, $j\in\{1,\ldots,l\}$ and $k\in\mathbb{N}$, are mutually independent, with
\begin{equation*}
\mathbb{E}[X_{j,k}]=0
\quad\text{and}\quad
\operatorname{Var}(X_{j,k})=\sigma_j^2,
\end{equation*}
where the centring assumption is not restrictive in the intended insurance application. In particular, when $X_{j,k}$ represents a future cash flow adjusted by the corresponding reserve, it is naturally centred. The aggregated portfolios before and after cancellation are given by
\begin{equation*}
S_n^\mathrm{pre}
\coloneqq
\sum_{j=1}^l
{}\underbrace{\sum_{k=1}^{n_j} X_{j,k}}_{\eqqcolon\, S_{j,n}^{\mathrm{pre}}}
\quad\text{and}\quad
S_n^\mathrm{post}
\coloneqq
\sum_{j=1}^l
\underbrace{\sum_{k=1}^{\tilde n_j} X_{j,k}}_{\eqqcolon\, S_{j,n}^{\mathrm{post}}},
\end{equation*}
where $\tilde n_j = (1-\theta_j)n_j + o(n)$ and the $o(n)$ term again accounts for rounding. In particular, by independence, the variance of the aggregated portfolio and its covariance with the loss of segment $j$ are given by
\begin{equation*}
\operatorname{Var}(S_n^\mathrm{pre}) =  n \langle w, \sigma^2 \rangle + o(n)
\quad\text{and}\quad
\operatorname{Cov}(S_{j,n}^{\mathrm{pre}},S_n^\mathrm{pre}) = n  w_j \sigma_j^2 + o(n),
\end{equation*}
where $w \coloneqq (w_1,\dots,w_l)$ and $\sigma^2 \coloneqq (\sigma_1^2,\dots,\sigma_l^2)$. Analogously, for the portfolio after cancellation, define $\tilde w_j\coloneqq(1-\theta_j)w_j$ and $\tilde w\coloneqq(\tilde w_1,\ldots,\tilde w_l)$, and conclude
\begin{equation*}
\operatorname{Var}(S_n^\mathrm{post}) =  n \langle \tilde w, \sigma^2 \rangle + o(n)
\quad\text{and}\quad
\operatorname{Cov}(S_{j,n}^{\mathrm{post}},S_n^\mathrm{post}) = n  \tilde w_j \sigma_j^2 + o(n).
\end{equation*}
To quantify the contribution of the individual segments to the aggregated portfolio risk, we use a standard-deviation allocation. This is particularly convenient in the present setting, as the normal approximation of the portfolio is determined by its variance. We define the relative risk contribution of segment $j$ to the standard deviation of the aggregate portfolio by
\begin{align*}
\mathrm{RC}_j^{\mathrm{pre}}
&\coloneqq 
\frac{
\operatorname{Cov}(S_{j,n}^{\mathrm{pre}},S_n^{\mathrm{pre}})
}{
\operatorname{Var}(S_n^{\mathrm{pre}})
} = 
\frac{w_j\sigma_j^2}
{\langle w,\sigma^2\rangle} + o(1),
\\ 
\mathrm{RC}_j^{\mathrm{post}}
&\coloneqq 
\frac{
\operatorname{Cov}(S_{j,n}^{\mathrm{post}},S_n^{\mathrm{post}})
}{
\operatorname{Var}(S_n^{\mathrm{post}})
} = 
\frac{\tilde w_j\sigma_j^2}
{\langle \tilde w,\sigma^2\rangle} + o(1). 
\end{align*}
Note that $\sum_{j=1}^l \mathrm{RC}_j^\mathrm{pre} =1 $ and $\sum_{j=1}^l \mathrm{RC}_j^\mathrm{post} =1 $, as well as that $\mathrm{RC}_j^{\mathrm{pre}}$ and $\mathrm{RC}_j^{\mathrm{post}}$ represent the relative contribution of segment $j$ to the total portfolio risk under the standard-deviation allocation.

\begin{definition}[Segment-level change in individual risk contribution]
\label{def:::DeltaHeterogeneous}
For $j\in\{1,\ldots,l\}$, we define the cancellation effect attributed to segment $j$ by
\begin{equation*}
\Delta_j^\mathrm{aggr}
\coloneqq 
\rho\biggl(
\frac{\mathrm{RC}_j^{\mathrm{post}}}{\tilde n_j}S_n^{\mathrm{post}}
\biggr)
-
\rho\biggl(
\frac{\mathrm{RC}_j^{\mathrm{pre}}}{n_j}S_n^{\mathrm{pre}}
\biggr).
\end{equation*}
\end{definition}
The quantity $\Delta_j^\mathrm{aggr}$ refers to the risk contribution of segment $j$ within the aggregated portfolio and should not be confused with the risk associated with the standalone segment losses $X_{j,n}^{\mathrm{pre}}$ and $X_{j,n}^{\mathrm{post}}$, which we denote by
\begin{multline*}
  \Delta_j \coloneqq  
  \rho\Bigl(\frac{1}{\tilde n_j} \sum_{k=1}^{\tilde n_j}X_{j,k}\Bigr)- \rho\Bigl(\frac{1}{ n_j}\sum_{k=1}^{n_j}X_{j,k}\Bigr) =
  \Bigl(\frac{1}{\tilde n_j^{1/2}} - \frac{1}{n_j^{1/2}} \Bigr) \sigma\rho(\Phi) + o\Bigl(\frac{1}{\tilde n_j^{1/2}}\Bigr) \\
  =
  \Bigl(\frac{1}{\tilde w_j} - \frac{1}{w_j}\Bigr) \frac{\sigma\rho(\Phi)}{n^{1/2}} + o\Bigl(\frac{1}{n^{1/2}}\Bigr) 
  =
  \frac{1-(1-\theta_j)^{1/2}}{(1-\theta_j)^{1/2}} \frac{\sigma \rho(\Phi)}{n_j^{1/2}} + o\Bigl(\frac{1}{n^{1/2}}\Bigr) .
\end{multline*}
In particular, for the risk contribution of segment $j$ within the aggregated portfolio, by positive homogeneity,
\begin{equation*}
\Delta_j^\mathrm{aggr} 
=
\frac{n^{1/2}\mathrm{RC}_j^\mathrm{post}}{\tilde n_j}
\rho\biggl[
\sum_{i=1}^l
\frac{\tilde n_i^{1/2}}{n^{1/2}}\biggl(
\frac{1}{\tilde n_i^{1/2}}\sum_{k=1}^{\tilde n_i} X_{i,k}\biggr)
\biggr]
-
\frac{n^{1/2}\mathrm{RC}_j^\mathrm{pre}}{ n_j}
\rho\biggl[
\sum_{i=1}^l
\frac{n_i^{1/2}}{n^{1/2}}\biggl(
\frac{1}{n_i^{1/2}}\sum_{k=1}^{n_i} X_{i,k}\biggr)
\biggr] ,
\end{equation*}
and we aim to derive the asymptotics of this quantity by the CLT in Wasserstein distance.

\pagebreak[2]
\begin{lemma}[Asymptotics for proportional cancellation in a heterogeneous portfolio]
\label{lem:HeterogeneousProportionalCancellation}

Let\/ $\rho$ be a law-invariant positive homogeneous risk measure that is Lipschitz continuous on\/ $(\mathcal{M}_1^p(\mathbb{R}),W_p)$ for some\/ $p \in [1,\infty)$. Let\/ $X_{j,k}$ be centred and mutually independent random variables with standard deviation\/ $\sigma_j$ and finite\/ $p$-th moment for\/ $j \in \{1,\dots,l\}$ and $k \in \mathbb{N}$. Then 
\begin{equation}
\label{eq:HeterogeneousProportionalCancellation}
\Delta_j^\mathrm{aggr} = 
\frac{\sigma_j^2}{(n\langle w,\sigma^2 \rangle)^{1/2}} 
\frac{1-(1-\bar\theta_\sigma)^{1/2}}{(1-\bar\theta_\sigma)^{1/2}}\rho(\Phi) 
+ o\Bigl(\frac{1}{n^{1/2}}\Bigr),
\end{equation}
where
\begin{equation*}
\bar\theta_\sigma \coloneqq \sum_{j=1}^l \pi_j \theta_j, 
\quad\text{with }\ \pi_j \coloneqq \frac{w_j \sigma_j^2}{\langle w,\sigma^2\rangle}.
\end{equation*}
\end{lemma}

As a consistency check, consider the special case $l=1$. Then $w_1=1$ and $\bar\theta_\sigma=\theta_1$, so \cref{lem:HeterogeneousProportionalCancellation} implies \cref{thm:::AsymptoticProportionalLp}, since in the homogeneous setting we can assume without loss of generality that the random variables are centred.
The Value-at-Risk does not fall within the continuity assumptions of the preceding result and therefore requires a separate argument. Nevertheless, the same asymptotic expression can be established in this case, as stated in the following corollary.

\pagebreak[2]
\begin{corollary}[Asymptotics in heterogeneous portfolio for VaR]
\label{cor:HeterogeneousProportionalCancellationVaR}
Let\/ $X_{j,k}$ be centred and mutually independent random variables with standard deviation\/ $\sigma_j$ and finite\/ $p$-th moment for\/ $j \in \{1,\dots,l\}$ and $k \in \mathbb{N}$. Then, for $\bar\theta_\sigma$ defined as in \cref{lem:HeterogeneousProportionalCancellation},
\begin{equation}
\label{eq:HeterogeneousProportionalCancellationVaR}
\Delta_{j,\alpha}^\mathrm{aggr}
=
\frac{\sigma_j^2}{(n\langle w,\sigma^2\rangle)^{1/2}}
\frac{1-(1-\bar\theta_\sigma)^{1/2}}
{(1-\bar\theta_\sigma)^{1/2}}
\Phi^{-1}(\alpha)
+
o\Bigl(\frac{1}{n^{1/2}}\Bigr).
\end{equation}
\end{corollary}

\pagebreak[2]
\begin{proof}
The proof follows the same argument as that of
\cref{lem:HeterogeneousProportionalCancellation}. The only difference is
the approximation of the risk measure. For each $i \in \{1,\dots,l\}$, the CLT gives
\begin{equation*}
\frac{1}{\tilde n_i^{1/2}}
\sum_{k=1}^{\tilde n_i}X_{i,k}
\xrightarrow{d} Z_i,
\qquad Z_i\sim\mathcal N(0,\sigma_i^2).
\end{equation*}
Since the segment losses are independent, their componentwise convergence implies joint convergence. Together with $\tilde n_i/n\to\tilde w_i$ this yields
\begin{equation*}
\sum_{i=1}^l
\frac{\tilde n_i^{1/2}}{n^{1/2}}
\biggl(
\frac{1}{\tilde n_i^{1/2}}
\sum_{k=1}^{\tilde n_i}X_{i,k}
\biggr)
\xrightarrow{d}
\sum_{i=1}^l\tilde w_i^{1/2}Z_i
\sim
\mathcal N(0,\langle\tilde w,\sigma^2\rangle).
\end{equation*}
Since the limiting distribution is continuous and strictly increasing,
the corresponding quantiles converge, and
\[
\operatorname{VaR}_\alpha\biggl(
\sum_{i=1}^l
\frac{\tilde n_i^{1/2}}{n^{1/2}}
\biggl(
\frac{1}{\tilde n_i^{1/2}}
\sum_{k=1}^{\tilde n_i}X_{i,k}
\biggr)
\biggr)
=
\langle\tilde w,\sigma^2\rangle^{1/2}
\Phi^{-1}(\alpha)+o(1).
\]
The corresponding result before cancellation follows analogously.
The remainder of the proof is identical to that of
\cref{lem:HeterogeneousProportionalCancellation}.
\end{proof}

\pagebreak[2]
\begin{proof}[Proof of \cref{lem:HeterogeneousProportionalCancellation}]
For random variables $X_1,\dots,X_l$ and $\tilde X_1,\dots,\tilde X_l$, each mutually independent, $W_p(X_1+\dots + X_l, \tilde X_1 + \dots + \tilde X_l) \le \sum_{i=1}^l W_p(X_i,\tilde X_i)$. Consider $Z_i \sim \mathcal{N}(0,\sigma_i^2)$ mutually independent, then
\begin{equation*}
W_p\Biggl( 
\sum_{i=1}^l \frac{\tilde n_i^{1/2}}{n^{1/2}}
\biggl(\frac{1}{\tilde n_i^{1/2}} \sum_{k=1}^{\tilde n_i} X_{i,k}\biggr),
\sum_{i=1}^l \frac{\tilde n_i^{1/2}}{n^{1/2}} Z_i
\Biggr) 
\le 
\sum_{i=1}^l \frac{\tilde n_i^{1/2}}{n^{1/2}}
{}\underbrace{W_p\biggl(
\frac{1}{\tilde n_i^{1/2}} \sum_{k=1}^{\tilde n_i} X_{i,k},
Z_i
\biggr)}_{=\,o(1)} = o(1)
\end{equation*}
as well as 
\begin{equation*}
W_p\Biggl( 
\sum_{i=1}^l \frac{n_i^{1/2}}{n^{1/2}}
\biggl(\frac{1}{n_i^{1/2}} \sum_{k=1}^{n_i} X_{i,k}\biggr),
\sum_{i=1}^l \frac{n_i^{1/2}}{n^{1/2}} Z_i
\Biggr) 
\le 
\sum_{i=1}^l \frac{ n_i^{1/2}}{n^{1/2}}
{}\underbrace{W_p\biggl(
\frac{1}{n_i^{1/2}} \sum_{k=1}^{n_i} X_{i,k},
Z_i
\biggr)}_{=\,o(1)} = o(1),
\end{equation*}
because $\tilde n_i/n = O(1)$ and $n_i/n = O(1)$.
Since $\sum_{i=1}^l (\tilde n_i/n)^{1/2}Z_i \sim \mathcal{N}(0,n^{-1}\sum_{i=1}^l  \tilde n_i \sigma_i^2)$ and $\sum_{i=1}^l (n_i/n)^{1/2}Z_i \sim \mathcal{N}(0,n^{-1}\sum_{i=1}^l  n_i \sigma_i^2)$, we obtain, similarly to the proof of \cref{thm:::AsymptoticProportionalLp} in the homogeneous case,
\begin{equation*}
\Delta_j^\mathrm{aggr} =
\Biggl(
\frac{\mathrm{RC}_j^\mathrm{post}}{\tilde n_j^{1/2}}
\biggl(\sum_{i=1}^l \frac{\tilde n_i}{\tilde n_j} \sigma_i^2\biggr)^{1/2} -
\frac{\mathrm{RC}_j^\mathrm{pre}}{n_j^{1/2}}
\biggl(\sum_{i=1}^l \frac{n_i}{n_j} \sigma_i^2\biggr)^{1/2}
\Biggr) \rho(\Phi) + o\Bigl(\frac{1}{n^{1/2}}\Bigr).
\end{equation*}
Note that $\mathrm{RC}_j^\mathrm{post}/ \tilde n_j^{1/2} = \frac{w_j^{1/2}\sigma_j^2}{\langle w,\sigma^2\rangle n^{1/2}} + o(n^{-1/2})$ and $\tilde n_i/ \tilde n_j = \tilde w_i / \tilde w_j + o(1)$, which gives
\begin{equation*}
\Delta_j^\mathrm{aggr}  
=
\biggl( 
\frac{\sigma_j^2}{\langle\tilde w,\sigma^2\rangle^{1/2} n^{1/2}} 
-
\frac{\sigma_j^2}{\langle w,\sigma^2\rangle^{1/2} n^{1/2}} 
\biggr)
\rho(\Phi) + o\Bigl(\frac{1}{n^{1/2}}\Bigr) . 
\end{equation*}
Use $\langle \tilde w,\sigma^2 \rangle = (1-\bar\theta_\sigma)\langle w,\sigma^2\rangle$ to conclude \eqref{eq:HeterogeneousProportionalCancellation}.
\end{proof}

\pagebreak[2]
The asymptotic expression in \eqref{eq:HeterogeneousProportionalCancellation} consists of two distinct effects. First, the cancellation effect is determined by the cancellation rate of the aggregate portfolio, represented by the variance-weighted average $\bar\theta_\sigma$. Second, aggregation provides a diversification effect through the factor $\sigma_j^2/(\langle w,\sigma^2\rangle)^{1/2}$.
Thus, the effect of aggregation on a particular segment is not determined by its own cancellation rate alone. 
The reduction due to diversification can be offset when the effective cancellation rate $\bar\theta_\sigma$ of the aggregate portfolio exceeds the segment-specific cancellation rate $\theta_j$.

\begin{example}[Aggregation may increase cancellation cost for individual segments]
\label{exm:::FailureAggregation}

Consider a portfolio consisting of two segments with identical weights ($w_1=w_2=0.5$) and identical variance ($\sigma_1=\sigma_2$), but different cancellation rates
\begin{equation*}
\theta_1=0.01
\quad\text{and}\quad
\theta_2=0.10.
\end{equation*}
Hence the effective cancellation rate in the aggregated portfolio is
\begin{equation*}
\bar\theta_\sigma
=
\frac{\theta_1+\theta_2}{2}
=
0.055,
\end{equation*}
although only $1\%$ of the contracts in segment~$1$ are cancelled. For comparison, if segment~$1$ were considered as a standalone portfolio of size $w_jn$, its corresponding leading-order cancellation effect would be
\begin{equation*}
\Delta_j
=
\frac{\sigma_j}{(w_jn)^{1/2}}
\frac{1-(1-\theta_j)^{1/2}}
{(1-\theta_j)^{1/2}}
\rho(\Phi)
+
o\Bigl(\frac{1}{n^{1/2}}\Bigr).
\end{equation*}
Consequently,
\begin{equation*}
\frac{\Delta_1^\mathrm{aggr}}{\Delta_1}
\longrightarrow
\frac{\sigma_1\sqrt{w_1}}
{\sqrt{\langle w,\sigma^2\rangle}}
\frac{(1-\bar\theta_\sigma)^{-1/2}-1}{(1-\theta_1)^{-1/2}-1}
=
\frac{1}{\sqrt{2}}
\frac{\sqrt{200/189}-1}{\sqrt{100/99}-1}
\approx 4.03.
\end{equation*}
Hence, despite the diversification benefit from aggregating the two segments, the cancellation effect attributed to segment $1$ is approximately four times larger in the aggregate portfolio than when the segment is considered in isolation, since the high cancellation rate of segment $2$ offsets the diversification benefit for segment $1$. This leads also to a cancellation fee roughly four times higher for segment $1$.
\end{example}

\pagebreak[2]
The example illustrates that the diversification effect does not necessarily dominate the effect of heterogeneous cancellation rates. A general upper bound for the cancellation effect follows from
\begin{equation*}
\langle w,\sigma^2\rangle \geq w_j\sigma_j^2,
\end{equation*}
which yields
\begin{equation*}
\Delta_j^\mathrm{aggr}
\leq
\frac{\sigma_j}{(w_j n)^{1/2}}
\frac{1-(1-\bar\theta_\sigma)^{1/2}}
{(1-\bar\theta_\sigma)^{1/2}}
\rho(\Phi)
+
o\Bigl(\frac{1}{n^{1/2}}\Bigr).
\end{equation*}
This bound corresponds to the cancellation effect that would arise if the segment were considered in isolation but subject to the aggregate cancellation rate $\bar\theta_\sigma$.
Since the function $(0,1) \ni u \mapsto f(u) \coloneqq (1-(1-u)^{1/2})/(1-u)^{1/2}$ is increasing, the bound is no larger than the corresponding standalone cancellation effect whenever $\theta_j\geq\bar\theta_\sigma$. Hence, if the cancellation rate $\theta_j$ of segment $j$ is at least as large as the effective cancellation rate of the aggregate portfolio $\bar\theta_\sigma$, aggregation reduces the change in individual risk contribution. 

The effect of aggregation becomes particularly transparent when the cancellation rates are identical across all segments ($\theta_1 = \dots = \theta_l$). Then $\bar\theta_\sigma=\theta_j$, and
\begin{equation*}
\Delta_j^\mathrm{aggr}
=
\frac{\sigma_j^2}{(n\langle w,\sigma^2\rangle)^{1/2}}
\frac{1-(1-\theta_j)^{1/2}}
{(1-\theta_j)^{1/2}}
\rho(\Phi)
+
o\Bigl(\frac{1}{n^{1/2}}\Bigr).
\end{equation*}
which yields
\begin{equation*}
\frac{\Delta_j^\mathrm{aggr}}
{\Delta_j}
\longrightarrow
\frac{ w_j^{1/2} \sigma_j}
{\langle w,\sigma^2\rangle^{1/2}} = \pi_j^{1/2}
\leq 1.
\end{equation*}
Thus, when cancellation rates are homogeneous across segments, aggregation reduces the cancellation effect attributed to each individual segment relative to its standalone effect. For equal segment variances $\sigma_j^2$, this reduction factor simplifies to $w_j^{1/2}$. In contrast, when cancellation rates differ across segments, this diversification effect need not hold uniformly across all segments, as the effective cancellation rate $\bar\theta_\sigma$ may exceed the cancellation rate of a particular segment (see \cref{exm:::FailureAggregation}).

\pagebreak[2]
\subsection{Extension to stochastic cancellation}
\label{sec:::ExtensionStochasticCancellation}

The proportional cancellation regime assumes that the fraction of cancelled policies is deterministic. We next extend this framework to a stochastic cancellation mechanism, where each policy is cancelled independently with a fixed probability. In this setting, the effective cancellation proportion becomes a random quantity that fluctuates around its deterministic limit.
More precisely, let $(I_{j,k})_{k\in\mathbb{N}}$, $j\in\{1,\dots,l\}$, be mutually independent Bernoulli random variables with $I_{j,k}\sim\mathrm{Bern}(\theta_j)$, indicating whether policy $k$ in segment $j$ cancels. Assume further that the family $(I_{j,k})_{j\in\{1,\dots,l\},k\in\mathbb{N}}$ is independent of the family of losses $(X_{j,k})_{j\in\{1,\dots,l\},k\in\mathbb{N}}$. 

As in the setting of the heterogeneous portfolio, suppose that the number of policies in segment $j\in\{1,\dots,l\}$ is $n_j=w_jn+o(n)$.
For fixed numbers of remaining policies $(\tilde n_1,\dots,\tilde n_l)$, let
\begin{equation*}
\Delta^{\mathrm{aggr}}(\tilde n_1,\dots,\tilde n_l) =
\left(\begin{array}{c}
\Delta_1^{\mathrm{aggr}}(\tilde n_1,\dots,\tilde n_l) \\
\vdots \\
\Delta_l^{\mathrm{aggr}}(\tilde n_1,\dots,\tilde n_l)
\end{array}\right)
\end{equation*}
denote the aggregate risk contribution difference defined in \cref{def:::DeltaHeterogeneous}. We then obtain the stochastic counterpart by replacing the deterministic numbers of remaining policies by their random values
\begin{equation*}
\tilde N_j
\coloneqq 
\sum_{k=1}^{n_j}(1-I_{j,k}),
\qquad j \in \{1,\dots,l\}.
\end{equation*}
Thus, we define
\begin{equation*}
\Delta^{\mathrm{aggr},\mathrm{stoch}} =
\Delta^{\mathrm{aggr}}(\tilde N_1,\dots,\tilde N_l).
\end{equation*}
This definition is useful because, conditionally on the cancellation indicators, the stochastic cancellation problem reduces to the deterministic setting considered in the previous subsection. The crucial assumption here is the independence between the cancellation indicators and the losses. Thus, the realisation of the cancellation mechanism provides information about the number of remaining policies, but does not provide any information about their associated losses. Conditional on the realised cancellation indicators, the remaining losses therefore retain the same distributional properties as in the deterministic setting.

\begin{lemma}[Stochastic extension of proportional cancellation]
\label{thm:::StochasticExtensionOfProportionalCancellation}

Let $\rho$ be a law-invariant positive homogeneous risk measure that is Lipschitz continuous on $(\mathcal{M}_1^p(\mathbb{R}),W_p)$ for some $p \in [1,\infty)$. Let $X_{j,k}$ be centred and mutually independent random variables with standard deviation $\sigma_j$ and finite $p$-th moment for $j\in \{1,\dots,l\}$ and $k\in\mathbb{N}$ with independent cancellation indicator $I_{j,k}\sim \mathrm{Bern}(\theta_{j})$. Then
\begin{equation*}
\Delta_j^{\mathrm{aggr},\mathrm{stoch}} 
\overset{\mathrm{a.s.}}{=}
\frac{\sigma_j^2}{(n\langle w, \sigma^2 \rangle)^{1/2}}
\frac{1-(1-\bar\theta_\sigma)^{1/2}}{(1-\bar\theta_\sigma)^{1/2}} \rho(\Phi) + 
o\Bigl(\frac{1}{n^{1/2}}\Bigr),
\end{equation*}
where, similar to \cref{lem:HeterogeneousProportionalCancellation},
\begin{equation*}
\bar\theta_\sigma \coloneqq \sum_{j=1}^l \pi_j \theta_j, 
\quad\text{with }\ \pi_j \coloneqq \frac{w_j \sigma_j^2}{\langle w,\sigma^2\rangle}.
\end{equation*}
\end{lemma}

\begin{proof}
Define $\Theta_{j,n} = \frac{1}{n_j}\sum_{k=1}^{n_j} I_{j,k}$. By the strong law of large numbers $\Theta_{j,n} \to \theta_j$ almost surely as $n \to \infty$. Hence, as $n \to \infty$,
\begin{equation*}
\tilde N_j = (1-\Theta_{j,n}) n_j 
\overset{\mathrm{a.s.}}{=}
(1-\theta_j) n_j + o(n_j) = \tilde w_j n + o(n),
\end{equation*}
where $\tilde w_j = (1-\theta_j)w_j$.
Thus, for almost every realisation $\omega \in \Omega$ of the cancellation indicators, the realised vector
\begin{equation*}
(\tilde N_1(\omega),\ldots,\tilde N_l(\omega))
\end{equation*}
satisfies the assumptions of \cref{lem:HeterogeneousProportionalCancellation}.
Since the cancellation indicators are independent of the losses, conditioning on their realisation does not alter the distribution of the losses in the remaining portfolio. We may therefore apply \cref{lem:HeterogeneousProportionalCancellation} pointwise to the realised vector of remaining policy numbers, which yields the stated result.
\end{proof}

\pagebreak[2]
\section{Numerical results}
\label{sec:::NumericalResults}

\subsection{Calculation of the cancellation cost}
\label{sec:CalculationOfTheCancellationCost}

While a detailed data-driven study of the risk-based component of the cancellation cost is beyond this work, this section provides tools for such calculations and presents some exemplary numerical values, which are not meant to be ready-to-use.
We consider two representative life insurance products, namely term life and pure endowment insurance (see the subsequent subsections). The calculations are based on the Austrian mortality table (see \cite[Life table for Austria, 2020--2022, males and females combined]{StatistikAustria2023LifeTables}) and are intended to assess the practical relevance of cancellation risk under realistic parameter choices. 
In particular, applying \cref{lem:::AsymptoticProportionalVaR,thm:::AsymptoticProportionalLp} to the homogeneous portfolio setting and \cref{lem:HeterogeneousProportionalCancellation,cor:HeterogeneousProportionalCancellationVaR} to the aggregated portfolio setting, we consider the quantities
\begin{equation}
\label{eq:CCRhat}
\mathrm{\widehat{CCR}} = 
\frac{1-\theta}{\theta}
\Bigl(
\frac{\sigma }{n^{1/2}} \frac{1-(1-\theta)^{1/2}}{(1-\theta)^{1/2}}
\Bigr)\rho(\Phi)
\end{equation}
and
\begin{equation}
\label{eq:CCRaggrhat}
\mathrm{\widehat{CCR}}_j^\mathrm{aggr} = 
\frac{1-\theta_j}{\theta_j} 
\Bigl(
\frac{\sigma_j^2}{(n\langle w, \sigma^2 \rangle)^{1/2}}
\frac{1-(1-\bar\theta_\sigma)^{1/2}}{(1-\bar\theta_\sigma)^{1/2}} 
\Bigr)\rho(\Phi),
\end{equation}
which we use as approximations of the cancellation cost in the
respective settings.
To evaluate these approximations, we need to determine the variance of the underlying individual loss.  To this end, we use Hattendorff's theorem, which provides a convenient way to calculate the variance of the future loss of a life insurance portfolio. The theorem in its classical form can be found, e.g., in \cite[Section 6.7]{Gerber1997}. For completeness, we state a simple extension to insurance products modelled by a Markov chain (see \cite{Norberg1992}). While the additional generality is not required for the examples considered below, the result provides a useful framework for assessing the variance of other insurance products modelled by a Markov chain (e.g.\ morbidity insurance with recovery).

\pagebreak[2]
\begin{theorem}[Hattendorff's theorem]
\label{thm:Hattendorff}
Let $(J_t)_{t \in \{0,\ldots,T\}}$ be a time-inhomogeneous Markov chain with
finite state space $\mathcal J$ and transition probabilities $p_{ij}(t) =
\mathbb P[J_{t+1}=j\mid J_t=i]$ for $i,j \in \mathcal{J}$ and $t \in \{0,\dots,T-1\}$.
Suppose that the insurance contract generates payments according to
the following scheme:
\begin{itemize}
    \item $b_i(t)$ is the payment at the beginning of period $t$
    conditional on $J_t=i$, for $t \in \{0,\ldots,T\}$;
    \item $b_{ij}(t)$ is the payment at the end of period $t$
    conditional on the transition $J_t=i$ to $J_{t+1}=j$,
    for $t \in \{0,\ldots,T-1\}$.
\end{itemize}
Let $v \in (0,1]$ denote the one-period discount factor. For
$t \in \{0,\ldots,T\}$, define the prospective present value of future cash flows
\begin{equation*}
L_t
\coloneqq
\sum_{k=t}^{T}v^{k-t}b_{J_k}(k)
+
\sum_{k=t}^{T-1}v^{k+1-t}b_{J_kJ_{k+1}}(k),
\end{equation*}
and the prospective reserve, before the payment at the beginning
of period $(t,t+1)$, by $V_t(i) = \mathbb{E}[L_t  \mid J_t = i]$.
Define
\begin{equation*}
\Lambda_{k}
\coloneqq
b_{J_k}(k) + v[b_{J_kJ_{k+1}}(k)+V_{k+1}(J_{k+1})]-V_k(J_k).
\end{equation*}
Then $\mathbb E[\Lambda_{t+1}\mid J_t]=0$ for $t \in \{0,\dots,T-1\}$ and $L_t-\mathbb{E}[L_t \mid J_t] = \sum_{k=t}^{T-1} v^{k-t}\Lambda_k$.
Consequently,
\begin{equation}
\label{eq:Hattendorff}
\operatorname{Var}(L_t\mid J_t)
=
\sum_{k=t}^{T-1}
v^{2(k-t)}
\mathbb E[\Lambda_k^2\mid J_t].
\end{equation}
\end{theorem}

\pagebreak[2]
The theorem follows from the standard martingale-difference argument underlying Hattendorff's theorem. For completeness, we provide the details in \cref{sec:ProofHattendorffTheorem}, adapted to the Markov-chain setting and the cash-flow convention used in the theorem and in the following examples (see also \cite{Norberg1992} for a proof of the general result).
With regard to the timing of cash flows, we define the prospective reserve $V_t$ immediately before the cash flow at time $t \in \{0,\dots T\}$. This convention is particularly convenient when considering cancellation at time $t$, since the corresponding premium payment is no longer made upon cancellation. Conditional on the current state $J_t$, the centred prospective loss is therefore given by $X_t \coloneqq L_t - V_t(J_t)$ with $\mathbb{E}[X_t\mid J_t]=0$. Thus, $X_t$ represents the remaining risk of the contract after accounting for the prospective reserve. The quantity $\Lambda_k$ can be interpreted as the one-period risk of the insurance contract over the period $(k,k+1)$.

In particular, \eqref{eq:Hattendorff} leads to a backward recursion for the conditional variance, which is particularly useful for practical calculations. For $i,j\in\mathcal J$, define
\begin{equation*}
\Lambda_{ij}(k)
= 
b_i(k) + v\left[b_{ij}(k)+V_{k+1}(j)\right]-V_k(i),
\end{equation*}
where  $\Lambda_k = \Lambda_{J_k,J_{k+1}}(k)$. Then Hattendorff's theorem implies that the variance can be calculated recursively by
\begin{equation}
\label{eq:RecursionVariance}
\operatorname{Var}(L_k\mid J_k=i)
=
\sum_{j\in\mathcal J}p_{ij}(k)
\bigl[
\Lambda_{ij}(k)^2
+
v^2\operatorname{Var}(L_{k+1}\mid J_{k+1}=j)
\bigr]
\end{equation}
and
\begin{equation*}
V_k(i) = 
b_i(k) + 
v \sum_{j \in \mathcal J} p_{ij}(k) [b_{ij}(k) + V_{k+1}(j)]
\end{equation*}
with terminal condition $\mathrm{Var}(L_T \mid J_T=i) = 0$ and $V_T(i) = b_i(T)$ for $i \in \mathcal{J}$.
This provides a convenient backward recursion for the numerical calculation of the variance (see, e.g., \cite[Theorem 5.4.1]{Koller2000} for a related recursion).

\subsection{Term life insurance}

We first consider a term life insurance contract with maturity $T=20$ and death benefit $B=100{,}000$. The insured is assumed to be aged $x=40$ at inception, and mortality is modelled according to the Austrian mortality table. 
A premium $P$ is paid at the beginning of each period while the insured is alive, whereas the death benefit $B$ is paid at the end of the period in which the insured dies. 

\begin{table}[ht!]
    \centering
    
\begin{tabular}{ccrrrr}
\toprule
$k$ & $x+k$ & $q_{x+k}$ & $V_k(a)$ & $\operatorname{Var}(L_k\mid J_k=a)$ & $\sqrt{\operatorname{Var}(L_k\mid J_k=a)}$ \\
\midrule
0 & 40 & 0.0009 & 0.00 & 288,421,935.35 & 16,982.99 \\
1 & 41 & 0.0010 & 145.92 & 291,515,701.70 & 17,073.83 \\
2 & 42 & 0.0010 & 287.45 & 294,044,535.72 & 17,147.73 \\
3 & 43 & 0.0011 & 423.63 & 295,901,194.18 & 17,201.78 \\
4 & 44 & 0.0012 & 553.37 & 296,965,782.06 & 17,232.70 \\
5 & 45 & 0.0014 & 675.47 & 297,105,042.61 & 17,236.74 \\
6 & 46 & 0.0015 & 788.65 & 296,171,567.06 & 17,209.64 \\
7 & 47 & 0.0016 & 891.47 & 294,002,473.36 & 17,146.50 \\
8 & 48 & 0.0018 & 982.27 & 290,406,229.99 & 17,041.31 \\
9 & 49 & 0.0020 & 1,058.99 & 285,147,123.95 & 16,886.30 \\
10 & 50 & 0.0022 & 1,119.15 & 277,941,827.68 & 16,671.59 \\
11 & 51 & 0.0024 & 1,159.84 & 268,456,136.68 & 16,384.63 \\
12 & 52 & 0.0027 & 1,177.72 & 256,301,577.30 & 16,009.42 \\
13 & 53 & 0.0030 & 1,169.11 & 241,043,459.49 & 15,525.57 \\
14 & 54 & 0.0033 & 1,130.16 & 222,208,725.15 & 14,906.67 \\
15 & 55 & 0.0037 & 1,056.77 & 199,281,354.05 & 14,116.70 \\
16 & 56 & 0.0041 & 944.64 & 171,696,325.22 & 13,103.29 \\
17 & 57 & 0.0046 & 789.18 & 138,829,284.76 & 11,782.58 \\
18 & 58 & 0.0051 & 584.86 & 99,927,946.56 & 9,996.40 \\
19 & 59 & 0.0057 & 324.68 & 54,042,876.81 & 7,351.39 \\
20 & 60 &  & 0.00 & 0.00 & 0.00 \\
\bottomrule
\end{tabular}

    \caption{Numerical results for the prospective reserve $V_k(a)$ and the conditional variance of the future loss $\operatorname{Var}(L_k \mid J_k = a)$ at times $k\in\{0,\ldots,T\}$ for the term life insurance contract, obtained using the backward recursion in \eqref{eq:RecursionTermLife}.}
    \label{tab:TermLifeResults}
\end{table}

\pagebreak[2]
Within the Markov-chain setup introduced in the preceding subsection, the term life insurance contract can be represented by the two states alive $a$ and dead $d$. The only nonzero cash flows are~$b_a(t)=-P$ and~$b_{ad}(t)=B$ for~$t \in \{0,\dots,T-1\}$ with all other payments equal to zero. In particular, in~\eqref{eq:RecursionVariance} we only have to consider the terms $p_{ad}\Lambda_{ad}(k)^2 = q_{x+k}[-P+vB-V_k(a)]^2$ and
$p_{aa}\Lambda_{aa}(k)^2  = p_{x+k} [-P+vV_{k+1}(a)-V_k(a)]^2$ for $k \in \{0,\dots,T-1\}$. Since $qA+pC=0$ and $p+q=1$ imply $qA^2+pC^2 = pq(A-C)^2$,
\begin{equation*}
p_{ad}\Lambda_{ad}(k)^2 + p_{aa}\Lambda_{aa}(k)^2
= 
q_{x+k} p_{x+k} \, v^2 [B-V_{k+1}(a)]^2,
\end{equation*}
which gives, since $\operatorname{Var}(L_k \mid J_k = d)= 0$, the recursions
\begin{equation}
\label{eq:RecursionTermLife}
\operatorname{Var}(L_k \mid J_k = a) = 
v^2 \bigl( 
q_{x+k} p_{x+k} [B-V_{k+1}(a)]^2 +
p_{x+k}\operatorname{Var}(L_{k+1} \mid J_{k+1} = a)
\bigr)
\end{equation}
and $V_k(a) = -P + v [q_{x+k} B + p_{x+k} V_{k+1}(a) ]$
for $k \in \{0,\dots,T-1\}$, with terminal conditions $\operatorname{Var}(L_T \mid J_T=a)=0$ and $V_T(a)=0$.
We evaluate the recursions above using an annual effective interest rate of $i=2\%$, corresponding to a discount factor of $v=\frac{1}{1+i}\approx 0.9804$. 
The corresponding net annual premium is calculated as $P=229.70$. The resulting prospective reserve and conditional variance are reported in~\cref{tab:TermLifeResults}.

\begin{table}[ht!]
    \centering
    \begin{tabular}{rrrrrrrr}
\toprule
\multicolumn{2}{c}{} & \multicolumn{2}{c}{$\theta=1\%$} & \multicolumn{2}{c}{$\theta=5\%$} & \multicolumn{2}{c}{$\theta=10\%$} \\
$k$ & $V_k(a)$ & $n_k$ & $\operatorname{\widehat{CCR}}_k$ & $n_k$ & $\operatorname{\widehat{CCR}}_k$ & $n_k$ & $\operatorname{\widehat{CCR}}_k$ \\
\cmidrule(lr){1-2}
\cmidrule(lr){3-4}
\cmidrule(lr){5-6}
\cmidrule(lr){7-8}
\midrule
1 & 145.92 & 50,000 & 88.59 & 50,000 & 87.68 & 50,000 & 86.48 \\
2 & 287.45 & 49,500 & 89.42 & 47,500 & 90.34 & 45,000 & 91.55 \\
3 & 423.63 & 49,005 & 90.16 & 45,125 & 92.98 & 40,500 & 96.81 \\
4 & 553.37 & 48,515 & 90.78 & 42,869 & 95.57 & 36,450 & 102.23 \\
5 & 675.47 & 48,030 & 91.25 & 40,725 & 98.08 & 32,805 & 107.78 \\
6 & 788.65 & 47,550 & 91.57 & 38,689 & 100.47 & 29,525 & 113.43 \\
7 & 891.47 & 47,074 & 91.69 & 36,755 & 102.70 & 26,572 & 119.13 \\
8 & 982.27 & 46,603 & 91.59 & 34,917 & 104.72 & 23,915 & 124.80 \\
9 & 1,058.99 & 46,137 & 91.21 & 33,171 & 106.46 & 21,523 & 130.36 \\
10 & 1,119.15 & 45,676 & 90.51 & 31,512 & 107.84 & 19,371 & 135.66 \\
11 & 1,159.84 & 45,219 & 89.40 & 29,937 & 108.74 & 17,434 & 140.54 \\
12 & 1,177.72 & 44,767 & 87.79 & 28,440 & 109.01 & 15,691 & 144.75 \\
13 & 1,169.11 & 44,319 & 85.57 & 27,018 & 108.46 & 14,121 & 147.97 \\
14 & 1,130.16 & 43,876 & 82.57 & 25,667 & 106.84 & 12,709 & 149.75 \\
15 & 1,056.77 & 43,437 & 78.59 & 24,384 & 103.81 & 11,438 & 149.49 \\
16 & 944.64 & 43,003 & 73.31 & 23,165 & 98.86 & 10,295 & 146.26 \\
17 & 789.18 & 42,573 & 66.26 & 22,006 & 91.20 & 9,265 & 138.63 \\
18 & 584.86 & 42,147 & 56.50 & 20,906 & 79.39 & 8,339 & 123.98 \\
19 & 324.68 & 41,726 & 41.76 & 19,861 & 59.90 & 7,505 & 96.11 \\
\bottomrule
\end{tabular}
    \caption{Cancellation costs $\operatorname{\widehat{CCR}}_k$ at time $k \in \{1,\dots,19\}$ under proportional cancellation for a new-business cohort of $n=50{,}000$ initial term life insurance contracts, where $\operatorname{\widehat{CCR}}_k$ is calculated according to \eqref{eq:CCRhat} for the $n_k$ contracts remaining in the portfolio at time $k$ and variance $\sigma_k^2 = \operatorname{Var}(L_k\mid J_k=a)$.}
    \label{tab:TermLifeCCR}
\end{table}

\pagebreak[2]
We use the conditional variances obtained above to evaluate the corresponding cancellation costs. To illustrate the impact of cancellations, we consider a constant annual cancellation probability $\theta\in\{1\%,5\%,10\%\}$
over the term of the contract.
As a reference risk measure, we use $\operatorname{VaR}_{0.99}$. For the standard normal distribution, $\operatorname{VaR}_{0.99}(\Phi)=\Phi^{-1}(0.99)\approx2.3263$. The numerical results for other admissible risk measures differ only through the multiplicative factor $\rho(\Phi)$ and therefore exhibit the same dependence on portfolio size, cancellation probability, and contract duration. We therefore restrict the numerical illustration to $\operatorname{VaR}_{0.99}$.
We consider initial portfolios of size $n=50{,}000$ consisting of homogeneous term life insurance contracts and interpret the portfolio as a new-business cohort. Risk is assumed to be pooled within each cohort. Under independent cancellations with constant annual cancellation probability $\theta$, the expected size of the remaining cohort after $k \in \{1,\dots,T\}$ years before any cancellation is $n_k = n(1-\theta)^{k-1}$.
This allows us to assess the cancellation cost for different cohort sizes and cancellation probabilities in \cref{tab:TermLifeCCR}.

\begin{table}[ht]
    \centering
    \begin{tabular}{rrrrrrrr}
\toprule
\multicolumn{2}{c}{} & \multicolumn{2}{c}{$\theta=1\%$} & \multicolumn{2}{c}{$\theta=5\%$} & \multicolumn{2}{c}{$\theta=10\%$} \\
$k$ & $V_k(a)$ & $n_k$ & $\operatorname{\widehat{CCR}}_k^\mathrm{aggr}$ & $n_k$ & $\operatorname{\widehat{CCR}}_k^\mathrm{aggr}$ & $n_k$ & $\operatorname{\widehat{CCR}}_k^\mathrm{aggr}$ \\
\cmidrule(lr){1-2}
\cmidrule(lr){3-4}
\cmidrule(lr){5-6}
\cmidrule(lr){7-8}
\midrule
1 & 145.92 & 50,000 & 23.23 & 50,000 & 26.47 & 50,000 & 30.54 \\
2 & 287.45 & 49,500 & 23.43 & 47,500 & 26.70 & 45,000 & 30.80 \\
3 & 423.63 & 49,005 & 23.57 & 45,125 & 26.87 & 40,500 & 31.00 \\
4 & 553.37 & 48,515 & 23.66 & 42,869 & 26.96 & 36,450 & 31.11 \\
5 & 675.47 & 48,030 & 23.67 & 40,725 & 26.98 & 32,805 & 31.12 \\
6 & 788.65 & 47,550 & 23.60 & 38,689 & 26.89 & 29,525 & 31.03 \\
7 & 891.47 & 47,074 & 23.42 & 36,755 & 26.69 & 26,572 & 30.80 \\
8 & 982.27 & 46,603 & 23.14 & 34,917 & 26.37 & 23,915 & 30.42 \\
9 & 1,058.99 & 46,137 & 22.72 & 33,171 & 25.89 & 21,523 & 29.87 \\
10 & 1,119.15 & 45,676 & 22.14 & 31,512 & 25.24 & 19,371 & 29.12 \\
11 & 1,159.84 & 45,219 & 21.39 & 29,937 & 24.37 & 17,434 & 28.12 \\
12 & 1,177.72 & 44,767 & 20.42 & 28,440 & 23.27 & 15,691 & 26.85 \\
13 & 1,169.11 & 44,319 & 19.20 & 27,018 & 21.89 & 14,121 & 25.25 \\
14 & 1,130.16 & 43,876 & 17.70 & 25,667 & 20.18 & 12,709 & 23.28 \\
15 & 1,056.77 & 43,437 & 15.88 & 24,384 & 18.09 & 11,438 & 20.88 \\
16 & 944.64 & 43,003 & 13.68 & 23,165 & 15.59 & 10,295 & 17.99 \\
17 & 789.18 & 42,573 & 11.06 & 22,006 & 12.60 & 9,265 & 14.54 \\
18 & 584.86 & 42,147 & 7.96 & 20,906 & 9.07 & 8,339 & 10.47 \\
19 & 324.68 & 41,726 & 4.31 & 19,861 & 4.91 & 7,505 & 5.66 \\
\bottomrule
\end{tabular}
    \caption{Cancellation costs $\operatorname{\widehat{CCR}}_k^{\mathrm{aggr}}$ at time $k \in \{1,\dots,19\}$ under proportional cancellation for an aggregated portfolio built from annual cohorts of $n=50{,}000$ new term life insurance contracts, where $\operatorname{\widehat{CCR}}_k^{\mathrm{aggr}}$ is calculated according to \eqref{eq:CCRaggrhat}, with $n_k$ and $\sigma_k^2=\operatorname{Var}(L_k\mid J_k=a)$ defined as in the non-aggregated case and portfolio weights $w_j = n_j / \sum_{j=1}^{19} n_j$ determined from the distribution of contracts in the total portfolio after several years of new business.}
    \label{tab:TermLifeCCRAgr}
\end{table}

\pagebreak[4]
In \cref{tab:TermLifeCCR}, we consider a new-business cohort consisting of homogeneous term life insurance contracts issued at the same time and assume that risk is pooled only within this cohort. We now extend the analysis to an insurer that has been selling the same product over an extended period and therefore holds an existing portfolio of contracts with different remaining maturities. We consider a portfolio built up over the preceding $20$ years, with a new cohort of $50{,}000$ term life insurance contracts with the same characteristics as above issued at the beginning of each year. For simplicity, cancellations are assumed to occur only at the end of each year. At the valuation date, the portfolio therefore consists of $20$ cohorts with remaining maturities ranging from $0$ to $19$ years. The cohort with remaining maturity $0$ has reached maturity and is no longer part of the in-force portfolio.
We evaluate the cancellation cost of the resulting portfolio using the same constant annual cancellation rates $\theta\in\{1\%,5\%,10\%\}$,  representing low, moderate, and high scenarios motivated by life-insurance lapse rates in the EIOPA Financial Stability Reports \cite{EIOPA2025FSR,EIOPA2026FSR}. The resulting cancellation costs are reported in \cref{tab:TermLifeCCRAgr}.

\begin{figure}[ht]
    \centering
    \includegraphics[width=\textwidth]{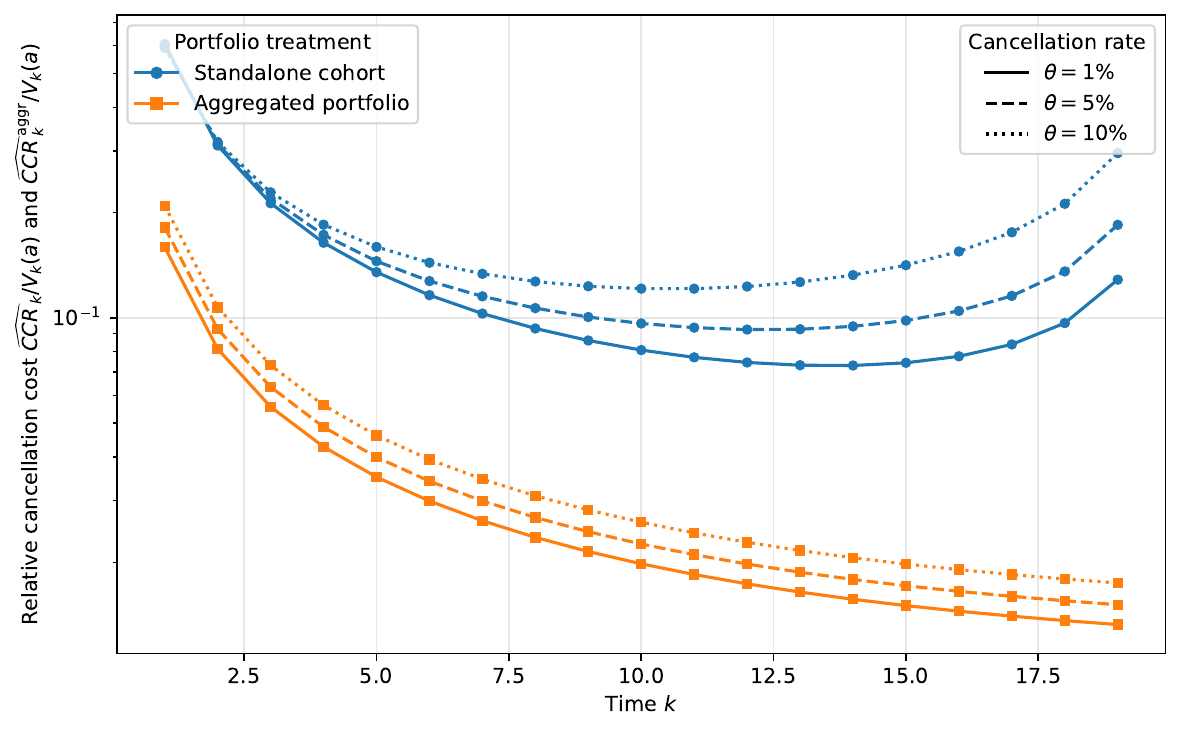}
    \caption{Relative cancellation costs for term life insurance under proportional cancellation, comparing a standalone new-business cohort (see \cref{tab:TermLifeCCR}) with an aggregated portfolio (see \cref{tab:TermLifeCCRAgr}). The cancellation costs are expressed relative to the prospective reserve.}
    \label{fig:relative_CCR_termlife_aggregated}
\end{figure}

\pagebreak[4]
To illustrate the effect of portfolio aggregation, we compare the relative cancellation costs of the aggregated portfolio with those of the corresponding standalone cohorts. Since the cancellation cost is most naturally assessed relative to the corresponding reserve in practice, we normalise the cancellation cost by the contract value $V_k(a)$, making the
relative cancellation costs independent of the contract size $B$. The resulting relative cancellation costs for the different cancellation probabilities are shown in \cref{fig:relative_CCR_termlife_aggregated}.
The figure shows that aggregating cohorts with different remaining maturities consistently reduces the relative cancellation cost, with the reduction being particularly pronounced for cohorts with shorter remaining maturities. The relative cancellation cost varies substantially over the lifetime of the contracts, while higher cancellation probabilities lead to higher cancellation costs. Overall, the figure illustrates how cancellation costs depend on both the cancellation probability and whether contracts are considered as a standalone cohort or as part of a diversified portfolio.

\subsection{Pure endowment}
We next consider a pure endowment insurance contract with maturity $T=20$ and benefit $B=100{,}000$. The insured is assumed to be aged $x=40$ at inception, and mortality is modelled according to the Austrian life table. A premium $P$ is paid at the beginning of each period while the insured is alive, whereas the benefit $B$ is paid at maturity if the insured survives to time $T$.

\pagebreak[2]
Within the Markov-chain setup introduced in the preceding subsection, the pure endowment insurance contract can again be represented by the two states alive~$a$ and dead~$d$. The only nonzero cash flows before maturity are the premium payments, while the benefit $B$ is paid at maturity if the insured is alive. Thus, for $t\in\{0,\ldots,T-1\}$, we have $b_a(t)=-P$, with all transition-dependent payments equal to zero, and the terminal payment is given by $b_a(T)=B$ and $b_d(T)=0$. Hence
\begin{equation}
\label{eq:RecursionPureEndowment}
\operatorname{Var}(L_k \mid J_k = a) = 
v^2\bigl(q_{x+k}p_{x+k} V_{k+1}(a)^2 + p_{x+k} \operatorname{Var}(L_{k+1} \mid J_{k+1} = a)\bigr)
\end{equation}
and $V_k(a) = -P + v\,p_{x+k}V_{k+1}(a)$ for $k \in \{0,\dots,T-1\}$, with terminal value $V_T(a)=B$ and $\operatorname{Var}(L_T \mid J_T = a) = 0$. Here,  we have used the same identity as in the term life insurance example to simplify the conditional variance.
For the numerical calculations, as in the term life example, we assume an annual effective interest rate of $i=2\%$, yielding the discount factor $v=(1+i)^{-1}\approx0.9804$. 
Using these assumptions, the corresponding net annual premium is $P=3,892.79$. The resulting prospective reserves and conditional variances are summarised in \cref{tab:EdowmentResults}.

\begin{table}[ht!]
    \centering
    
\begin{tabular}{ccrrrr}
\toprule
$k$ & $x+k$ & $q_{x+k}$ & $V_k(a)$ & $\operatorname{Var}(L_k\mid J_k=a)$ & $\sqrt{\operatorname{Var}(L_k\mid J_k=a)}$ \\
\midrule
0 & 40 & 0.0009 & 0.00 & 122,423,296.48 & 11,064.51 \\
1 & 41 & 0.0010 & 3,974.16 & 127,468,042.62 & 11,290.17 \\
2 & 42 & 0.0010 & 8,032.00 & 132,683,218.68 & 11,518.82 \\
3 & 43 & 0.0011 & 12,175.98 & 138,033,120.75 & 11,748.75 \\
4 & 44 & 0.0012 & 16,408.79 & 143,467,049.03 & 11,977.77 \\
5 & 45 & 0.0014 & 20,733.34 & 148,915,138.07 & 12,203.08 \\
6 & 46 & 0.0015 & 25,152.79 & 154,283,326.75 & 12,421.08 \\
7 & 47 & 0.0016 & 29,670.57 & 159,447,311.34 & 12,627.24 \\
8 & 48 & 0.0018 & 34,290.49 & 164,244,059.37 & 12,815.77 \\
9 & 49 & 0.0020 & 39,016.82 & 168,459,517.65 & 12,979.20 \\
10 & 50 & 0.0022 & 43,854.43 & 171,813,422.86 & 13,107.76 \\
11 & 51 & 0.0024 & 48,808.82 & 173,940,403.46 & 13,188.65 \\
12 & 52 & 0.0027 & 53,886.29 & 174,366,826.42 & 13,204.80 \\
13 & 53 & 0.0030 & 59,093.98 & 172,486,867.08 & 13,133.43 \\
14 & 54 & 0.0033 & 64,439.84 & 167,537,109.08 & 12,943.61 \\
15 & 55 & 0.0037 & 69,932.72 & 158,564,289.30 & 12,592.23 \\
16 & 56 & 0.0041 & 75,582.48 & 144,387,169.96 & 12,016.12 \\
17 & 57 & 0.0046 & 81,400.06 & 123,549,386.72 & 11,115.28 \\
18 & 58 & 0.0051 & 87,398.17 & 94,218,496.96 & 9,706.62 \\
19 & 59 & 0.0057 & 93,592.05 & 54,042,876.81 & 7,351.39 \\
20 & 60 &  & 100,000.00 & 0.00 & 0.00 \\
\bottomrule
\end{tabular}

    \caption{Numerical results for the prospective reserve $V_k(a)$ and the conditional variance of the future loss $\operatorname{Var}(L_k \mid J_k = a)$ at times $k\in\{0,\ldots,T\}$ for the pure endowment insurance contract, obtained using the backward recursion in \eqref{eq:RecursionPureEndowment}.}
    \label{tab:EdowmentResults} 
\end{table}

\pagebreak[2]
We use the conditional variances obtained above to evaluate the corresponding cancellation costs. To illustrate the impact of cancellations, we consider again a constant annual cancellation rate $\theta\in\{1\%,5\%,10\%\}$ over the term of the contract (cf.\ \cite{EIOPA2025FSR,EIOPA2026FSR}). As a reference risk measure, we continue to use $\operatorname{VaR}_{0.99}$. For the standard normal distribution, $\operatorname{VaR}_{0.99}(\Phi)=\Phi^{-1}(0.99)\approx2.3263$. 
We consider initial portfolios of size $n=50{,}000$ consisting of homogeneous pure endowment insurance contracts and interpret the portfolio as a new-business cohort. Risk is assumed to be pooled within each cohort. The size of the remaining cohort after $k\in\{1,\dots,T\}$ years before any cancellation is $n_k=n(1-\theta)^{k-1}$. This allows us to assess the cancellation cost for different cohort sizes and cancellation probabilities in \cref{tab:EndowmentCCR}.

\begin{table}[ht!]
    \centering
    \begin{tabular}{rrrrrrrr}
\toprule
\multicolumn{2}{c}{} & \multicolumn{2}{c}{$\theta=1\%$} & \multicolumn{2}{c}{$\theta=5\%$} & \multicolumn{2}{c}{$\theta=10\%$} \\
$k$ & $V_k(a)$ & $n_k$ & $\operatorname{\widehat{CCR}}_k$ & $n_k$ & $\operatorname{\widehat{CCR}}_k$ & $n_k$ & $\operatorname{\widehat{CCR}}_k$ \\
\cmidrule(lr){1-2}
\cmidrule(lr){3-4}
\cmidrule(lr){5-6}
\cmidrule(lr){7-8}
\midrule
1 & 3,974.16 & 50,000 & 58.58 & 50,000 & 57.98 & 50,000 & 57.18 \\
2 & 8,032.00 & 49,500 & 60.07 & 47,500 & 60.69 & 45,000 & 61.50 \\
3 & 12,175.98 & 49,005 & 61.58 & 45,125 & 63.51 & 40,500 & 66.12 \\
4 & 16,408.79 & 48,515 & 63.09 & 42,869 & 66.43 & 36,450 & 71.05 \\
5 & 20,733.34 & 48,030 & 64.60 & 40,725 & 69.43 & 32,805 & 76.31 \\
6 & 25,152.79 & 47,550 & 66.09 & 38,689 & 72.51 & 29,525 & 81.87 \\
7 & 29,670.57 & 47,074 & 67.53 & 36,755 & 75.63 & 26,572 & 87.73 \\
8 & 34,290.49 & 46,603 & 68.88 & 34,917 & 78.75 & 23,915 & 93.86 \\
9 & 39,016.82 & 46,137 & 70.11 & 33,171 & 81.83 & 21,523 & 100.20 \\
10 & 43,854.43 & 45,676 & 71.16 & 31,512 & 84.79 & 19,371 & 106.66 \\
11 & 48,808.82 & 45,219 & 71.96 & 29,937 & 87.53 & 17,434 & 113.12 \\
12 & 53,886.29 & 44,767 & 72.41 & 28,440 & 89.91 & 15,691 & 119.39 \\
13 & 59,093.98 & 44,319 & 72.38 & 27,018 & 91.75 & 14,121 & 125.17 \\
14 & 64,439.84 & 43,876 & 71.70 & 25,667 & 92.77 & 12,709 & 130.03 \\
15 & 69,932.72 & 43,437 & 70.10 & 24,384 & 92.60 & 11,438 & 133.34 \\
16 & 75,582.48 & 43,003 & 67.23 & 23,165 & 90.66 & 10,295 & 134.13 \\
17 & 81,400.06 & 42,573 & 62.50 & 22,006 & 86.04 & 9,265 & 130.78 \\
18 & 87,398.17 & 42,147 & 54.86 & 20,906 & 77.09 & 8,339 & 120.39 \\
19 & 93,592.05 & 41,726 & 41.76 & 19,861 & 59.90 & 7,505 & 96.11 \\
\bottomrule
\end{tabular}
    \caption{Cancellation costs $\operatorname{\widehat{CCR}}_k$ at time $k \in \{1,\dots,19\}$ under proportional cancellation for a new-business cohort of $n=50{,}000$ initial pure endowment insurance contracts, where $\operatorname{\widehat{CCR}}_k$ is calculated according to \eqref{eq:CCRhat} for the $n_k$ contracts remaining in the portfolio at time $k$ and variance $\sigma_k^2 = \operatorname{Var}(L_k\mid J_k=a)$.}
    \label{tab:EndowmentCCR}
\end{table}

\pagebreak[4]
We next consider an existing portfolio of pure endowment contracts built up over the preceding $20$ years, with a new cohort of $50{,}000$ contracts issued at the beginning of each year. Cancellations are assumed to occur only at the end of each year, resulting in $20$ cohorts with different remaining maturities at the valuation date. Using the same cancellation probabilities $\theta\in\{1\%,5\%,10\%\}$, we evaluate the cancellation costs within the resulting portfolio in \cref{tab:EndowmentCCRAgr}.

\begin{figure}[ht]
    \centering
    \includegraphics[width=\textwidth]{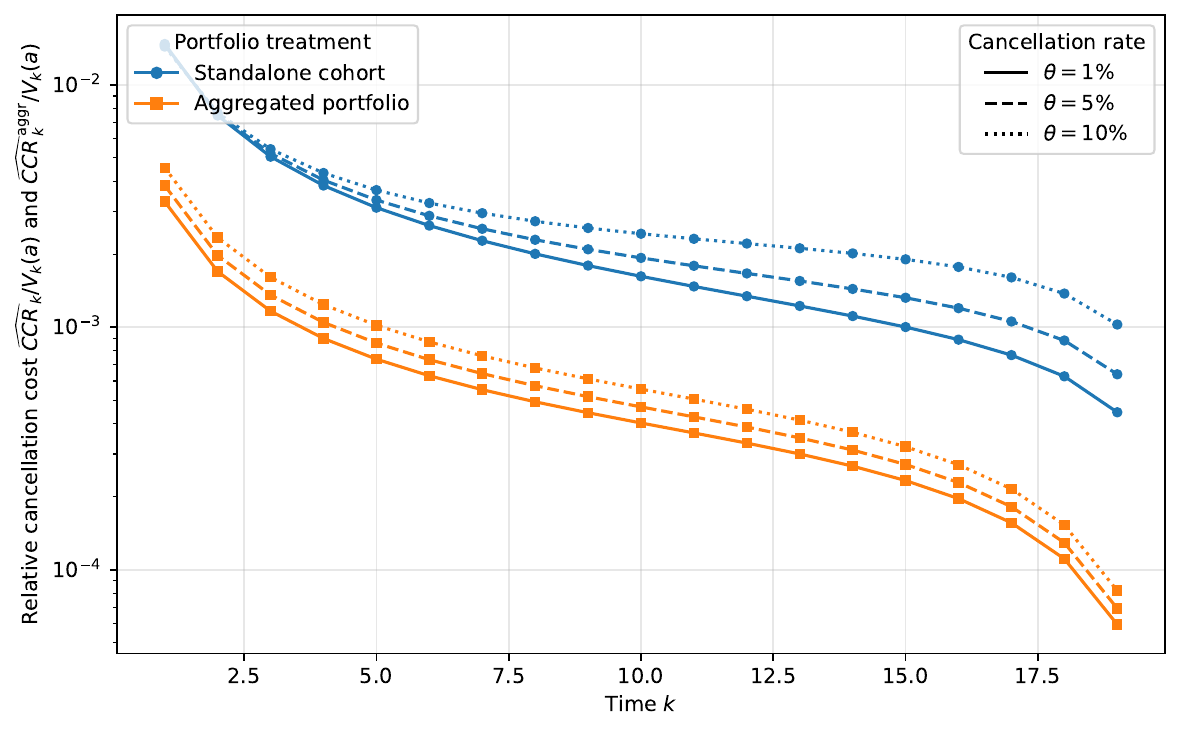}
    \caption{Relative cancellation costs for pure endowment insurance under proportional cancellation, comparing a standalone new-business cohort (see \cref{tab:EndowmentCCR}) with an aggregated portfolio (see \cref{tab:EndowmentCCRAgr}). The cancellation costs are expressed relative to the prospective reserve.}
    \label{fig:relative_CCR_endowment_aggregated}
\end{figure}

\pagebreak[2]
To illustrate the effect of portfolio aggregation, we compare the relative cancellation costs of the aggregated portfolio with those of the corresponding standalone cohorts. We normalise the cancellation cost by the contract value $V_k(a)$, which represents the value of the pure endowment contract at the time of cancellation. As a result, the relative cancellation cost is independent of the contract size $B$ as in the term life example. The resulting relative cancellation costs for the different cancellation probabilities are shown in \cref{fig:relative_CCR_endowment_aggregated}.

\begin{table}[ht]
    \centering
    \begin{tabular}{rrrrrrrr}
\toprule
\multicolumn{2}{c}{} & \multicolumn{2}{c}{$\theta=1\%$} & \multicolumn{2}{c}{$\theta=5\%$} & \multicolumn{2}{c}{$\theta=10\%$} \\
$k$ & $V_k(a)$ & $n_k$ & $\operatorname{\widehat{CCR}}_k^\mathrm{aggr}$ & $n_k$ & $\operatorname{\widehat{CCR}}_k^\mathrm{aggr}$ & $n_k$ & $\operatorname{\widehat{CCR}}_k^\mathrm{aggr}$ \\
\cmidrule(lr){1-2}
\cmidrule(lr){3-4}
\cmidrule(lr){5-6}
\cmidrule(lr){7-8}
\midrule
1 & 3,974.16 & 50,000 & 13.11 & 50,000 & 15.28 & 50,000 & 18.09 \\
2 & 8,032.00 & 49,500 & 13.65 & 47,500 & 15.90 & 45,000 & 18.83 \\
3 & 12,175.98 & 49,005 & 14.20 & 45,125 & 16.54 & 40,500 & 19.59 \\
4 & 16,408.79 & 48,515 & 14.76 & 42,869 & 17.20 & 36,450 & 20.36 \\
5 & 20,733.34 & 48,030 & 15.32 & 40,725 & 17.85 & 32,805 & 21.13 \\
6 & 25,152.79 & 47,550 & 15.87 & 38,689 & 18.49 & 29,525 & 21.89 \\
7 & 29,670.57 & 47,074 & 16.40 & 36,755 & 19.11 & 26,572 & 22.63 \\
8 & 34,290.49 & 46,603 & 16.90 & 34,917 & 19.69 & 23,915 & 23.31 \\
9 & 39,016.82 & 46,137 & 17.33 & 33,171 & 20.19 & 21,523 & 23.91 \\
10 & 43,854.43 & 45,676 & 17.68 & 31,512 & 20.59 & 19,371 & 24.38 \\
11 & 48,808.82 & 45,219 & 17.90 & 29,937 & 20.85 & 17,434 & 24.68 \\
12 & 53,886.29 & 44,767 & 17.94 & 28,440 & 20.90 & 15,691 & 24.74 \\
13 & 59,093.98 & 44,319 & 17.75 & 27,018 & 20.67 & 14,121 & 24.48 \\
14 & 64,439.84 & 43,876 & 17.24 & 25,667 & 20.08 & 12,709 & 23.78 \\
15 & 69,932.72 & 43,437 & 16.31 & 24,384 & 19.00 & 11,438 & 22.50 \\
16 & 75,582.48 & 43,003 & 14.85 & 23,165 & 17.31 & 10,295 & 20.49 \\
17 & 81,400.06 & 42,573 & 12.71 & 22,006 & 14.81 & 9,265 & 17.53 \\
18 & 87,398.17 & 42,147 & 9.69 & 20,906 & 11.29 & 8,339 & 13.37 \\
19 & 93,592.05 & 41,726 & 5.56 & 19,861 & 6.48 & 7,505 & 7.67 \\
\bottomrule
\end{tabular}
    \caption{Cancellation costs $\operatorname{\widehat{CCR}}_k^{\mathrm{aggr}}$ at time $k \in \{1,\dots,19\}$ under proportional cancellation for an aggregated portfolio built from annual cohorts of $n=50{,}000$ new pure endowment insurance contracts, where $\operatorname{\widehat{CCR}}_k^{\mathrm{aggr}}$ is calculated according to \eqref{eq:CCRaggrhat}, with $n_k$ and $\sigma_k^2=\operatorname{Var}(L_k\mid J_k=a)$ defined as in the non-aggregated case and portfolio weights $w_j = n_j / \sum_{j=1}^{19} n_j$ determined from the distribution of contracts in the total portfolio after several years of new business.}
    \label{tab:EndowmentCCRAgr}
\end{table}

\pagebreak[4]
The figure shows that pooling cohorts with different remaining maturities again leads to consistently lower relative cancellation costs than considering the cohorts separately. For $\theta=5\%$, the relative cancellation cost of the aggregated portfolio is reduced to approximately $26\%$ of the corresponding standalone value at $k=1$, with this ratio decreasing to approximately $11\%$ at $k=19$. As in the term life insurance example, the relative cancellation cost decreases substantially over the lifetime of the contracts and increases with the cancellation probability.

\pagebreak[4]
\section{Conclusion}

The asymptotic results established in this paper provide a mathematical framework for quantifying the effect of policy cancellation on portfolio risk and understanding its dependence on portfolio size and cancellation rates.
The numerical analysis based on these results illustrates the magnitude of cancellation costs for life insurance contracts. We consider a reference new-business cohort of $n=50{,}000$ contracts. For pure endowments, the resulting relative cancellation costs are of the order of $0.01\%$ to $1\%$ of the reserve. For term life insurance, the corresponding relative cancellation costs are substantially larger, reaching values of the order of $1\%$ to $100\%$ of the reserve, depending on the contract duration. Pure endowments are arguably of greater practical relevance in this context, as they are primarily savings-oriented products with comparatively higher reserves, whereas term life insurance is primarily designed to provide risk protection rather than capital accumulation. Nevertheless, the numerical results demonstrate that the magnitude of cancellation costs can vary considerably across insurance products.
The Markov-chain formulation developed in this paper provides a flexible framework for extending the numerical analysis to other insurance products and payment structures. Moreover, portfolio size must be considered when interpreting the numerical results, as the cancellation cost per cancelling policy decreases with increasing portfolio size.

\pagebreak[2]
The analysis of an existing portfolio with cohorts of different remaining maturities further shows that portfolio aggregation can substantially reduce the relative cancellation cost compared with considering the cohorts separately. However, such a reduction is not guaranteed, in particular when cancellation rates differ across cohorts. Thus, the magnitude of the cancellation cost depends not only on the cancellation probability and portfolio size, but also on the composition of the underlying portfolio. The framework also allows for the analysis of risk sharing across different insurance products or, more generally, across different homogeneous portfolio segments, although this is not considered in the numerical examples. The extent to which such diversification reduces cancellation costs again depends on the characteristics of the individual segments.

\pagebreak[2]
The analysis does not account for potentially relevant features of cancellation. In the model considered here, the cancellation cost arises solely from the increase in portfolio risk resulting from the reduction in the number of contracts, without accounting for any additional dependence between cancellation behaviour and policyholder risk characteristics. Such dependence may affect both the magnitude and the structure of the cancellation cost. Accounting for such dependence would therefore provide a more comprehensive assessment of cancellation risk and represents a natural extension of the present analysis.

\pagebreak[2]
\section*{AI disclaimer}

During the preparation of this manuscript, the authors used AI-assisted tools to improve the language and readability of the text and to assist in identifying related research. All mathematical results and proofs were independently developed and verified by the authors, who take full responsibility for the content of the manuscript.

\pagebreak[4]
\bibliography{literature}

@article {Bo18,
    AUTHOR = {Bobkov, Sergey G.},
     TITLE = {Central limit theorem and {D}iophantine approximations},
   JOURNAL = {J. Theoret. Probab.},
  FJOURNAL = {Journal of Theoretical Probability},
    VOLUME = {31},
      YEAR = {2018},
    NUMBER = {4},
     PAGES = {2390--2411},
      ISSN = {0894-9840,1572-9230},
   MRCLASS = {60F05 (11J04 11J82)},
  MRNUMBER = {3866618},
MRREVIEWER = {Rita\ Giuliano Antonini},
       DOI = {10.1007/s10959-017-0770-4},
       URL = {https://doi.org/10.1007/s10959-017-0770-4},
}

@article {KoMc90,
    AUTHOR = {Kolassa, John E. and McCullagh, Peter},
     TITLE = {Edgeworth series for lattice distributions},
   JOURNAL = {Ann. Statist.},
  FJOURNAL = {The Annals of Statistics},
    VOLUME = {18},
      YEAR = {1990},
    NUMBER = {2},
     PAGES = {981--985},
      ISSN = {0090-5364,2168-8966},
   MRCLASS = {62E20 (60F05)},
  MRNUMBER = {1056348},
MRREVIEWER = {Stavros\ Kourouklis},
       DOI = {10.1214/aos/1176347637},
       URL = {https://doi.org/10.1214/aos/1176347637},
}

@article {DoFe23,
    AUTHOR = {Dolgopyat, Dmitry and Fernando, Kasun},
     TITLE = {An error term in the central limit theorem for sums of
              discrete random variables},
   JOURNAL = {Int. Math. Res. Not. IMRN},
  FJOURNAL = {International Mathematics Research Notices. IMRN},
      YEAR = {2023},
    NUMBER = {21},
     PAGES = {18664--18713},
      ISSN = {1073-7928,1687-0247},
   MRCLASS = {60F05 (11J13 37A17 37A50)},
  MRNUMBER = {4665633},
MRREVIEWER = {Vladimir\ I.\ Chebotarev},
       DOI = {10.1093/imrn/rnad088},
       URL = {https://doi.org/10.1093/imrn/rnad088},
}

@incollection {GovZ06,
    AUTHOR = {G\"otze, Friedrich and van Zwet, Willem R.},
     TITLE = {An expansion for a discrete non-lattice distribution},
 BOOKTITLE = {Frontiers in statistics},
     PAGES = {257--274},
 PUBLISHER = {Imp. Coll. Press, London},
      YEAR = {2006},
      ISBN = {1-86094-670-4},
   MRCLASS = {62E20},
  MRNUMBER = {2326005},
MRREVIEWER = {Qiying\ Wang},
       DOI = {10.1142/9781860948886\_0012},
       URL = {https://doi.org/10.1142/9781860948886_0012},
}

@article {HuLa19,
    AUTHOR = {Hu, Tian-You and Lau, Ka-Sing},
     TITLE = {Limiting behavior of infinite products scaled by {P}isot
              numbers},
   JOURNAL = {J. Fourier Anal. Appl.},
  FJOURNAL = {The Journal of Fourier Analysis and Applications},
    VOLUME = {25},
      YEAR = {2019},
    NUMBER = {4},
     PAGES = {1695--1707},
      ISSN = {1069-5869,1531-5851},
   MRCLASS = {42A38 (11B37 11R06)},
  MRNUMBER = {3977132},
MRREVIEWER = {P\'eter\ K\'orus},
       DOI = {10.1007/s00041-018-9638-y},
       URL = {https://doi.org/10.1007/s00041-018-9638-y},
}

@article {ChDeSt22,
    AUTHOR = {Chiarolla, Maria B. and De Angelis, Tiziano and Stabile,
              Gabriele},
     TITLE = {An analytical study of participating policies with minimum
              rate guarantee and surrender option},
   JOURNAL = {Finance Stoch.},
  FJOURNAL = {Finance and Stochastics},
    VOLUME = {26},
      YEAR = {2022},
    NUMBER = {2},
     PAGES = {173--216},
      ISSN = {0949-2984,1432-1122},
   MRCLASS = {91G80 (35R35 60G40 62P05)},
  MRNUMBER = {4397929},
MRREVIEWER = {Scott\ Robertson},
       DOI = {10.1007/s00780-022-00471-0},
       URL = {https://doi.org/10.1007/s00780-022-00471-0},
}

@article {DOSo16,
  AUTHOR = {D’Ortona, Nicolino Ettore and Staffa, Maria Sole},
  TITLE = {The theoretical surrender value in life insurance},
 JOURNAL = {Insurance Markets and Companies},
  YEAR = {2016},
  VOLUME ={7},
  NUMBER = {1},
  PAGES = {31-44},
}

@article {Ya13,
    AUTHOR = {Yakubovich, Semyon},
     TITLE = {On some {R}ajchman measures and equivalent {S}alem's problem},
   JOURNAL = {Commun. Math. Anal.},
  FJOURNAL = {Communications in Mathematical Analysis},
    VOLUME = {14},
      YEAR = {2013},
    NUMBER = {1},
     PAGES = {28--41},
      ISSN = {1938-9787},
   MRCLASS = {42B10 (28A12 33C10 44A15)},
  MRNUMBER = {3040879},
MRREVIEWER = {Lillian\ Beatrix\ Pierce},
}

@article {HuLa02,
    AUTHOR = {Hu, Tian-You and Lau, Ka-Sing},
     TITLE = {Fourier asymptotics of {C}antor type measures at infinity},
   JOURNAL = {Proc. Amer. Math. Soc.},
  FJOURNAL = {Proceedings of the American Mathematical Society},
    VOLUME = {130},
      YEAR = {2002},
    NUMBER = {9},
     PAGES = {2711--2717},
      ISSN = {0002-9939,1088-6826},
   MRCLASS = {42A38 (26A12 26A30)},
  MRNUMBER = {1900879},
MRREVIEWER = {Walter\ Trebels},
       DOI = {10.1090/S0002-9939-02-06398-0},
       URL = {https://doi.org/10.1090/S0002-9939-02-06398-0},
}

@article {BeDi17,
author = {Bellini, Fabio and Di Bernardino, Elena},
title = {Risk management with expectiles},
journal = {The {E}uropean Journal of Finance},
volume = {23},
number = {6},
pages = {487-506},
year = {2017},
}

@misc{DAV22,
type={Fachgrundsatz der Deutschen Aktuarvereinigung e. V.},
title={Hinweis: Stornoabzüge in der {Lebensversicherung}},
author={Deutsche Aktuarvereinigung},
address={K{\"o}ln},
day=27,
month=jan,
year=2022}

@book {Fel2,
    AUTHOR = {Feller, William},
     TITLE = {An introduction to probability theory and its applications. {Volume}~{II}},
   EDITION = {Second},
 PUBLISHER = {John Wiley \& Sons, Inc.}, 
   ADDRESS = {New York-London-Sydney},
      YEAR = {1971}
}

@article {BaMu06,
    AUTHOR = {B\"{a}uerle, Nicole and M\"{u}ller, Alfred},
     TITLE = {Stochastic orders and risk measures: consistency and bounds},
   JOURNAL = {Insurance Math. Econom.},
  FJOURNAL = {Insurance: Mathematics \& Economics},
    VOLUME = {38},
      YEAR = {2006},
    NUMBER = {1},
     PAGES = {132--148},
      ISSN = {0167-6687,1873-5959},
   MRCLASS = {60E15 (91B30)},
  MRNUMBER = {2197307},
MRREVIEWER = {Lorenzo\ Peccati},
       DOI = {10.1016/j.insmatheco.2005.08.003},
       URL = {https://doi.org/10.1016/j.insmatheco.2005.08.003},
}

@book {BhRa76,
    AUTHOR = {Bhattacharya, Rabindra N. and Ranga Rao, R.},
     TITLE = {Normal approximation and asymptotic expansions},
    SERIES = {Wiley Series in Probability and Mathematical Statistics},
 PUBLISHER = {John Wiley \& Sons, New York-London-Sydney},
      YEAR = {1976},
     PAGES = {xiv+274},
   MRCLASS = {60F05},
  MRNUMBER = {436272},
MRREVIEWER = {C.\ C.\ Heyde},
}

@book {Ka04,
    AUTHOR = {Katznelson, Yitzhak},
     TITLE = {An introduction to harmonic analysis},
   EDITION = {Third},
 PUBLISHER = {Cambridge University Press},
      YEAR = {2004},
     PAGES = {xviii+314},
      ISBN = {0-521-83829-0; 0-521-54359-2},
   MRCLASS = {43-01 (42-02 43-02)},
  MRNUMBER = {2039503},
       DOI = {10.1017/CBO9781139165372},
       URL = {https://doi.org/10.1017/CBO9781139165372},
}

@book {Hal92,
    AUTHOR = {Hall, Peter},
     TITLE = {The bootstrap and {E}dgeworth expansion},
 PUBLISHER = {Springer-Verlag, New York},
      YEAR = {1992}
}

@book {MuSt02,
    AUTHOR = {M\"{u}ller, Alfred and Stoyan, Dietrich},
     TITLE = {Comparison methods for stochastic models and risks},
    SERIES = {Wiley Series in Probability and Statistics},
 PUBLISHER = {John Wiley \& Sons, Ltd., Chichester},
      YEAR = {2002},
     PAGES = {xii+330},
      ISBN = {0-471-49446-1},
   MRCLASS = {60-02 (60E15 60K25 62E10 62P05)},
  MRNUMBER = {1889865},
MRREVIEWER = {Tomasz\ J.\ Kozubowski},
}

@article {ArVi86,
    AUTHOR = {Arnold, Barry C. and Villase\~nor, Jos\'e{} A.},
     TITLE = {Lorenz ordering of means and medians},
   JOURNAL = {Statist. Probab. Lett.},
  FJOURNAL = {Statistics \& Probability Letters},
    VOLUME = {4},
      YEAR = {1986},
    NUMBER = {1},
     PAGES = {47--49},
      ISSN = {0167-7152,1879-2103},
   MRCLASS = {62E99},
  MRNUMBER = {822726},
MRREVIEWER = {K.\ D.\ Ling},
       DOI = {10.1016/0167-7152(86)90039-8},
       URL = {https://doi.org/10.1016/0167-7152(86)90039-8},
}

@book {Villani2003,
    AUTHOR = {Villani, C\'edric},
     TITLE = {Topics in optimal transportation},
    SERIES = {Graduate Studies in Mathematics},
    VOLUME = {58},
 PUBLISHER = {American Mathematical Society, Providence, RI},
      YEAR = {2003},
     PAGES = {xvi+370},
      ISBN = {0-8218-3312-X},
   MRCLASS = {90-02 (28D05 35B65 35J60 49N90 49Q20 90B20)},
  MRNUMBER = {1964483},
       DOI = {10.1090/gsm/058},
       URL = {https://doi.org/10.1090/gsm/058},
}

@article {Whitney1987,
    AUTHOR = {Newey, Whitney K. and Powell, James L.},
     TITLE = {Asymmetric least squares estimation and testing},
   JOURNAL = {Econometrica},
  FJOURNAL = {Econometrica. Journal of the Econometric Society},
    VOLUME = {55},
      YEAR = {1987},
    NUMBER = {4},
     PAGES = {819--847},
      ISSN = {0012-9682,1468-0262},
   MRCLASS = {62J05 (62P20)},
  MRNUMBER = {906565},
       DOI = {10.2307/1911031},
       URL = {https://doi.org/10.2307/1911031},
}

@article {Rio2009,
    AUTHOR = {Rio, Emmanuel},
     TITLE = {Upper bounds for minimal distances in the central limit
              theorem},
   JOURNAL = {Ann. Inst. Henri Poincar\'e{} Probab. Stat.},
  FJOURNAL = {Annales de l'Institut Henri Poincar\'e{} Probabilit\'es et
              Statistiques},
    VOLUME = {45},
      YEAR = {2009},
    NUMBER = {3},
     PAGES = {802--817},
      ISSN = {0246-0203,1778-7017},
   MRCLASS = {60F05},
  MRNUMBER = {2548505},
MRREVIEWER = {Paul\ H.\ Bezandry},
       DOI = {10.1214/08-AIHP187},
       URL = {https://doi.org/10.1214/08-AIHP187},
}

@misc{Ha25,
      title={Non-uniform {E}dgeworth expansions for weakly dependent random variables and their applications}, 
      author={Yeor Hafouta},
      year={2025},
      eprint={2511.06414},
      archivePrefix={arXiv},
      note={arXiv:2511.06414 [\mbox{math.PR}]},
      url={https://arxiv.org/abs/2511.06414}, 
}

@misc{goulard2025,
      title={{Orlicz--Lorentz} premia and distortion {Haezendonck--Goovaerts} risk measures}, 
      author={Aline Goulard and Karl Grosse-Erdmann},
      year={2025},
      eprint={2512.03267},
      archivePrefix={arXiv},
      primaryClass={q-fin.RM},
      note={arXiv:2512.03267 [q\mbox{-}fin.RM]},
      url={https://arxiv.org/abs/2512.03267}, 
}

@article {Bobkov2018,
    AUTHOR = {Bobkov, Sergey G.},
     TITLE = {Berry-{E}sseen bounds and {E}dgeworth expansions in the
              central limit theorem for transport distances},
   JOURNAL = {Probab. Theory Related Fields},
  FJOURNAL = {Probability Theory and Related Fields},
    VOLUME = {170},
      YEAR = {2018},
    NUMBER = {1-2},
     PAGES = {229--262},
      ISSN = {0178-8051,1432-2064},
   MRCLASS = {60F05},
  MRNUMBER = {3748324},
MRREVIEWER = {Vladimir\ I.\ Chebotarev},
       DOI = {10.1007/s00440-017-0756-2},
       URL = {https://doi.org/10.1007/s00440-017-0756-2},
}

@article{Burkhart2018,
  author  = {Burkhart, Tobias},
  title   = {Surrender Risk in the Context of the Quantitative Assessment of Participating Life Insurance Contracts under {Solvency II}},
  journal = {Risks},
  volume  = {6},
  number  = {3},
  articleno = {66},
  year    = {2018},
  doi     = {10.3390/risks6030066}
}

@article{GoSm21,
Author = {Gottlieb, Daniel and Smetters, Kent},
Title = {Lapse-Based Insurance},
Journal = {American Economic Review},
Volume = {111},
Number = {8},
Year = {2021},
Month = {August},
Pages = {2377–2416},
DOI = {10.1257/aer.20160868},
}

@article{Hacariz2026,
	author = {Ha{\c c}arız, Oytun and Kleinow, Torsten and Macdonald, Angus S.},
	date = {2026/04/01},
	doi = {10.1007/s13385-025-00434-7},
	id = {Ha{\c c}arız2026},
	isbn = {2190-9741},
	journal = {European Actuarial Journal},
	number = {1},
	pages = {191--223},
	title = {Lapse-supported life insurance and adverse selection},
	url = {https://doi.org/10.1007/s13385-025-00434-7},
	volume = {16},
	year = {2026}}

@article {Chang2022,
    AUTHOR = {Chang, Hsiaoyin and Schmeiser, Hato},
     TITLE = {Life insurance surrender and liquidity risks},
   JOURNAL = {Quant. Finance},
  FJOURNAL = {Quantitative Finance},
    VOLUME = {22},
      YEAR = {2022},
    NUMBER = {4},
     PAGES = {761--776},
      ISSN = {1469-7688,1469-7696},
   MRCLASS = {91G05},
  MRNUMBER = {4416040},
       DOI = {10.1080/14697688.2021.1998586},
       URL = {https://doi.org/10.1080/14697688.2021.1998586},
}

@article {KainaRueschendorf2009,
    AUTHOR = {Kaina, M. and R\"uschendorf, L.},
     TITLE = {On convex risk measures on {$L^p$}-spaces},
   JOURNAL = {Math. Methods Oper. Res.},
  FJOURNAL = {Mathematical Methods of Operations Research},
    VOLUME = {69},
      YEAR = {2009},
    NUMBER = {3},
     PAGES = {475--495},
      ISSN = {1432-2994,1432-5217},
   MRCLASS = {60B11 (62P05 91B30)},
  MRNUMBER = {2507760},
MRREVIEWER = {Frank\ Riedel},
       DOI = {10.1007/s00186-008-0248-3},
       URL = {https://doi.org/10.1007/s00186-008-0248-3},
}

@article{Li05,
title={Changes in Pure Premium Policy-Values consequent upon variations in the Rate of Interest or the Rate of Mortality, or upon the introduction of the Rate of Discontinuance},
volume={39},
DOI={10.1017/S0020268100008441}, 
number={2}, 
journal={Journal of the Institute of Actuaries}, 
author={Lidstone, George James}, 
year={1905},
pages={209–252}}

@article {Artzner1999,
    AUTHOR = {Artzner, Philippe and Delbaen, Freddy and Eber, Jean-Marc and
              Heath, David},
     TITLE = {Coherent measures of risk},
   JOURNAL = {Math. Finance},
  FJOURNAL = {Mathematical Finance. An International Journal of Mathematics,
              Statistics and Financial Economics},
    VOLUME = {9},
      YEAR = {1999},
    NUMBER = {3},
     PAGES = {203--228},
      ISSN = {0960-1627,1467-9965},
   MRCLASS = {91B30 (91B28)},
  MRNUMBER = {1850791},
MRREVIEWER = {Yanhong\ Wu},
       DOI = {10.1111/1467-9965.00068},
       URL = {https://doi.org/10.1111/1467-9965.00068},
}

@book {Gerber1997,
    AUTHOR = {Gerber, Hans U.},
     TITLE = {Life insurance mathematics},
   EDITION = {Third},
      NOTE = {With exercises contributed by Samuel H. Cox,
              With a foreword by Hans B\"uhlmann,
              Translated from the German by Walther Neuhaus},
 PUBLISHER = {Springer-Verlag, Berlin; Association of Swiss Actuaries,
              Z\"urich},
      YEAR = {1997},
     PAGES = {xviii+217},
      ISBN = {3-540-62242-X},
   MRCLASS = {62P05},
  MRNUMBER = {1439045},
       DOI = {10.1007/978-3-662-03460-6},
       URL = {https://doi.org/10.1007/978-3-662-03460-6},
}

@misc{StatistikAustria2023LifeTables,
  author       = {{S}tatistik {A}ustria},
  title        = {Detailed general and balanced life tables 1868--71 to 2020--22},
  year         = {2023},
  institution  = {{S}tatistik {A}ustria},
  note         = {Sterbetafel 2020/2022 männlich und weiblich zusammen},
  url          = {https://www.statistik.at/en/statistics/population-and-society/population/demographic-indicators-and-tables/life-tables}
}

@article {Norberg1992,
    AUTHOR = {Norberg, Ragnar},
     TITLE = {Hattendorff's theorem and {T}hiele's differential equation
              generalized},
   JOURNAL = {Scand. Actuar. J.},
  FJOURNAL = {Scandinavian Actuarial Journal},
      YEAR = {1992},
    NUMBER = {1},
     PAGES = {2--14},
      ISSN = {0346-1238},
   MRCLASS = {62P05},
  MRNUMBER = {1193668},
       DOI = {10.1080/03461238.1992.10413894},
       URL = {https://doi.org/10.1080/03461238.1992.10413894},
}

@book{Koller2000,
publisher = {Springer},
booktitle = {Stochastische Modelle in der Lebensversicherung},
isbn = {3540660569},
year = {2000},
title = {Stochastische Modelle in der Lebensversicherung},
language = {ger},
address = {Berlin},
author = {Koller, Michael},
}

@book {Wheeden2015,
    AUTHOR = {Wheeden, Richard L. and Zygmund, Antoni},
     TITLE = {Measure and integral},
    SERIES = {Pure and Applied Mathematics (Boca Raton)},
   EDITION = {Second},
      NOTE = {An introduction to real analysis},
 PUBLISHER = {CRC Press, Boca Raton, FL},
      YEAR = {2015},
     PAGES = {xvii+514},
      ISBN = {978-1-4987-0289-8},
   MRCLASS = {28-01 (26-01 42-01)},
  MRNUMBER = {3381284},
}

@article {Johnson2005,
    AUTHOR = {Johnson, Oliver and Samworth, Richard},
     TITLE = {Central limit theorem and convergence to stable laws in
              {M}allows distance},
   JOURNAL = {Bernoulli},
  FJOURNAL = {Bernoulli. Official Journal of the Bernoulli Society for
              Mathematical Statistics and Probability},
    VOLUME = {11},
      YEAR = {2005},
    NUMBER = {5},
     PAGES = {829--845},
      ISSN = {1350-7265,1573-9759},
   MRCLASS = {60F05 (60E07)},
  MRNUMBER = {2172843},
MRREVIEWER = {Peter\ Kern},
       DOI = {10.3150/bj/1130077596},
       URL = {https://doi.org/10.3150/bj/1130077596},
}

@report{EIOPA2025FSR,
  author      = {{European Insurance and Occupational Pensions Authority}},
  title       = {Financial Stability Report},
  date        = {2025-06-19},
  year        = {2025},
  month       = jun,
  day         = 19,
  url         = {https://www.eiopa.europa.eu/publications/financial-stability-report-june-2026_en},
}

@report{EIOPA2026FSR,
  author      = {{European Insurance and Occupational Pensions Authority}},
  title       = {Financial Stability Report},
  date        = {2026-06-24},
  year        = {2026},
  month       = jun,
  day         = 24,
  url         = {https://www.eiopa.europa.eu/publications/financial-stability-report-june-2025_en},
}

\pagebreak
\appendix
\section{Hattendorff's theorem for Markov chains}
\label{sec:ProofHattendorffTheorem}

This appendix contains a self-contained proof of the Hattendorff theorem stated in \cref{thm:Hattendorff} in the Markov-chain setting considered in this paper. The argument is included for completeness, since the result is a standard consequence of the martingale-difference representation underlying Hattendorff's theorem, but we use a slightly more general formulation that accommodates time-inhomogeneous transition probabilities and the cash-flow convention adopted here.

\pagebreak[2]
\begin{proof}[Proof of \cref{thm:Hattendorff}]
The result follows from the standard argument underlying Hattendorff's theorem. We provide the details here since we formulate the result in the specific Markov chain setting and cash-flow convention considered in this paper, which is natural for our subsequent analysis. 
By the Markov property, the prospective reserve satisfies the discrete-time Thiele difference equation
\begin{equation*}
V_k(i)
= b_i(k) + v\sum_{j\in\mathcal J}p_{ij}(k) [b_{ij}(k)+V_{k+1}(j)],
\qquad i \in \mathcal{J},k\in\{0,\dots,T-1\},
\end{equation*}
where $V_T(i) = b_i(T)$. Consequently,
\begin{equation*}
\mathbb E[\Lambda_k\mid\mathcal F_k]
=
b_{J_k}(k) + 
v\,
\mathbb E\left[
b_{J_kJ_{k+1}}(k)
+V_{k+1}(J_{k+1})
\,\middle|\,\mathcal F_k
\right]
-
V_k(J_k)
=0.
\end{equation*} 
Thus $(\Lambda_k)_{k=0}^{T-1}$ is a martingale-difference sequence. By the definition of $\Lambda_k$
\begin{equation*}
v^{k-t}\Lambda_k
=
v^{k-t}
\bigl(
b_{J_k}(k) + v[b_{J_kJ_{k+1}}(k)+V_{k+1}(J_{k+1})] \bigr)
-v^{k-t}V_k(J_k)
,
\end{equation*}
and summing from $k=t$ to $T-1$ yields
\begin{multline*}
\sum_{k=t}^{T-1}v^{k-t}\Lambda_k
=
\sum_{k=t}^{T-1}
v^{k+1-t}b_{J_kJ_{k+1}}(k)
+
\sum_{k=t}^{T-1}
v^{k-t}b_{J_k}(k) \\
+
\underbrace{
\sum_{k=t}^{T-1}v^{k+1-t}V_{k+1}(J_{k+1})
-
\sum_{k=t}^{T-1}v^{k-t}V_k(J_k) }_{=\; v^{T-t}V_T(J_{k+1}) - V_t(J_t) \; = \; v^{T-t}b_{J_T}(T)- V_t(J_t)}.
\end{multline*}
Conclude
\begin{equation*}
\sum_{k=t}^{T-1}v^{k-t}\Lambda_k
=
L_t-V_t(J_t) = L_t - \mathbb{E}[L_t \mid J_t].
\end{equation*}
Finally, since $(\Lambda_k)_{k \in \{ 0,\dots T-1\}}$ is a martingale-difference sequence, for
$t\leq k<\ell\leq T-1$,
\begin{equation*}
\mathbb E[\Lambda_k\Lambda_\ell\mid J_t]
=
\mathbb E\left[
\Lambda_k
\mathbb E[\Lambda_\ell\mid\mathcal F_{\ell-1}]
\,\middle|\,J_t
\right]
=0.
\end{equation*}
Hence
\begin{equation*}
\operatorname{Var}(L_t\mid J_t)
=
\mathbb E\Bigl[
\Bigl(
\sum_{k=t}^{T-1}v^{k-t}\Lambda_k
\Bigr)^2
\,\Bigr|\,J_t
\Bigr]
=
\sum_{k=t}^{T-1}
v^{2(k-t)}
\mathbb E[\Lambda_k^2\mid J_t]. \qedhere
\end{equation*}
\end{proof}

\end{document}